\documentclass{aa}
\usepackage{graphicx}
\usepackage{natbib}
\bibpunct{(}{)}{;}{a}{}{,} 

\begin{document}

\titlerunning{LMC Self-lensing from a new perspective}

\title{LMC Self-lensing from a new perspective}

\authorrunning{L. Mancini et al.}

\author{L. Mancini\inst{1,2,4}, S. Calchi Novati\inst{1},
        Ph. Jetzer\inst{1}, G. Scarpetta\inst{2,3,4}}
\institute{Institut f\"{u}r Theoretische Physik der
           Universit\"{a}t Z\"{u}rich, CH-8057 Z\"{u}rich, Switzerland
           \and Dipartimento di Fisica ``E.R. Caianiello'',
           Universit\`{a} di Salerno, I-84081 Baronissi (SA), Italy
           \and International Institute for Advanced Scientific Studies,
           Vietri sul Mare (SA), Italy
           \and Istituto Nazionale di Fisica Nucleare, sez. Napoli, Italy}
\date{Received  / Accepted}

\abstract{
We present a new analysis on the issue of the location of the
observed microlensing events in direction of the Large Magellanic
Cloud (LMC). This is carried out starting from a recently drawn
coherent picture of the geometrical structure and dynamics of the
LMC disk and by considering different
configurations for the LMC bar. In this framework it clearly emerges that
the spatial distribution of the events observed so far shows a
near--far asymmetry. This turns out to be compatible with the
optical depth calculated for the LMC halo objects. In this
perspective, our main conclusion, supported by a statistical
analysis on the outcome of an evaluation of the microlensing rate,
is that self lensing can not account for all the observed events.
Finally we propose a general inequality to calculate quickly an
upper limit to the optical depth along a line of view through the
LMC center.

\keywords{gravitational lensing -- dark matter -- galaxies:
magellanic clouds}}

\maketitle

\section{Introduction}
The microlensing surveys towards the Large Magellanic Cloud (LMC)
\citep{alcock00a,lasserre00} have demonstrated the existence of
compact objects that act as gravitational lenses somewhere between
us and the LMC. In some cases the distance and the mass of the lenses
have been determined, thanks to the proper motion of the lens
observed by the Hubble Space Telescope (HST)
\citep{alcock01a,gould04,drake04} or to the additional information
carried by binary systems \citep{alcock00b,alcock01b}. However,
these are special cases since for most events only the duration
and the position on the sky plane have been measured. These
information are not enough to establish definitively if the detected
events are really caused by white dwarfs  or MACHOs in the halo of
the Milky Way (MW), or are due to stars or MACHOs  within the LMC
itself.

The survey of the MACHO team indicates a most probable
Galactic halo fraction of 20\%, with limits of 5\% to 50\% at the
95\% confidence level, assuming that all the events are due to
halo lenses. The preferred value for the lens mass is $\sim$ 0.4
M$_{\sun}$. This is consistent with the EROS survey results, that
are given however as an upper limit for the Galactic halo
fraction.

An interesting alternative is that of ``self lensing'',
where both source and lens belong to the luminous  part of the LMC as
suggested by \citet{sahu94} and \citet{wu94}. However, the initial
estimates of the optical depth and microlensing rate were lower
than the measured one \citep{gould95,alcock97a,alcock00a}. The
self--lensing explanation has been further analyzed going beyond
the hypothesis of a ``simple'' geometry for the LMC  with  disk
and bar coplanar, so that their relative distance would enhance
the optical depth and, therefore, the rate.  In the model of
\citet{zhao00} the disk and bar stars are on two distinct planes
with different inclinations, so that stars on the front plane
could lens those in the plane $\approx$ 1 kpc behind. Besides the
morphology, another aspect considered has been the dynamics of the
luminous components within the LMC.  \citet{aubourg99}, by using a
model which takes into account the correlation between the mass of
the lenses in the LMC and their velocity dispersion, have been able to
reproduce a self--lensing optical depth, event rate and event
duration distribution compatible with the observed ones. Yet,
objections to this model have been raised by different authors
\citep{gyuk00,alves00}, especially with respect to the adopted
distribution and velocity dispersion of the lensing stars, which
seem to be inconsistent with the observations.

The analysis of Jetzer et al.,  (2002, hereafter Paper I) has
shown that probably the observed events are distributed among
different  components (disk, spheroid and galactic halo, the LMC halo
and self--lensing). This means that the lenses do not belong all
to the same population and their astrophysical features can differ
deeply from one another.

In this paper we address once more the question of the presence of
a self--lensing component within the LMC itself. To this end a
correct knowledge of the structure and dynamics of the luminous
components (disk and bar) of the LMC is essential. Here we
take advantage of some recent studies of the LMC disk (see Sect.
\ref{sec:disk-morphology}), while we allow for different geometries
for the still poorly known bar component, to calculate the main
microlensing quantities. Moreover, with respect to Paper I, based
on the moment method \citep{derujula}, we perform instead a
statistical analysis starting from the differential rate of the
microlensing events.

The paper is organized as follows: in Sects. \ref{sec:morphology} and
\ref{sec:modelli} we discuss the LMC morphology and present the models
we use to describe the spatial density of the MW halo and of the
various components of the LMC. Sect. \ref{sec:mlpar} is devoted to the
calculation of the microlensing quantities, the optical depth and
the microlensing rate, as well as to a statistical analysis of the
self--lensing events. In Sect. \ref{sec:asimmetria} we discuss the
spatial asymmetry with respect to the line of nodes of the
observed microlensing events. An improved inequality for the
optical depth for self lensing by a stellar disk is derived in
Sect. \ref{sec:gould}. We conclude in Sect. \ref{sec:conclusioni} with a
summary of our results.

%
\section{Observational data}
\label{sec:morphology}
\subsection{LMC Disk morphology}
\label{sec:disk-morphology}
Recently, in a series of three interesting papers
\citep{marel01a,marel01b,marel02}, a new coherent picture of the
 geometrical structure and dynamics of the LMC disk has been
given. By using the data from two
near-infrared\footnote{Near-infrared surveys provide a very
accurate view, thanks to their insensitivity to the dust
absorption.} surveys, DENIS \citep{cioni00} and 2MASS
\citep{cutri00} stellar catalogs, the LMC geometry has been
constrained by carefully studying the star count map and the
characteristic apparent magnitudes of Asymptotic Giant Branch
(AGB) and Red Giant Branches (RGB) stars. The results have
been further improved by a detailed and sophisticated study of
the LMC line-of-sight velocity field, coupled with a
multi-parameter fit on the available velocities of 1041 carbon
stars. A basic assumption is that the carbon star population
is representative for the bulk of the LMC stars.

The first important conclusion is that the intrinsic shape of the
LMC disk is not circular, as assumed before, but
elliptical, with an intrinsic ellipticity $\epsilon^{\prime\prime}
= 0.312 \pm 0.007$. The inclination angle of the LMC disk plane is
$i = 34.7^{\circ} \pm 6^{\circ}$ and the line-of-nodes position
angle is $\Theta = 129.9^{\circ} \pm 6^{\circ}$. This value
is quite different from the  LMC disk major axis position angle,
$\mathrm{PA}_{\mathrm{maj}} = 189.3^{\circ}\pm1.4^{\circ}$,
corresponding to a position angle $\Phi =
202.7^{\circ}\pm1.9^{\circ}$ when measured in the equatorial plane
of the LMC disk, starting from the axis pointing towards the North. 
The radial number density profile along the major axis follows, to lowest
order, an exponential profile with an intrinsic scale length equal
to 1.54 kpc.

A second important conclusion is that the  center of mass (CM) of
the carbon stars is consistent with the center of the bar and with
the center of the outer isophotes of the LMC. As a consequence,
the idea of using the distribution of neutral gas as  good tracer
for the disk stars, that leads to an incorrect LMC model, must be
abandoned.  The obtained values of the right ascension
 and  declination  of the CM, given in
J$2000$, are $\alpha_{\mathrm{CM}} = 5^{\mathrm{h}} \,
27.6^{\mathrm{m}} \pm 3.9^{\mathrm{m}}$, $\delta_{\mathrm{CM}} =
-\, 69.87^{\circ} \pm 0.41^{\circ}$. The weighted mean of the
rotation velocity in the range 4--8.9 kpc, where the rotation
curve is approximately flat, is $V = 49.8 \pm 15.9$ km/s, about
$40\% $ lower than the previously inferred and accepted value.
Taking into account the asymmetric drift effect, the circular
velocity  has been corrected and estimated to be equal to
$V_{\mathrm{circ}} = 64.8 \pm 15.9$ $\mathrm{km}\,
\mathrm{s}^{-1}$. The line-of-sight velocity dispersion has an
average value $\sigma=20.2\pm0.5$ km $\mathrm{s}^{-1}$, and is
little dependent on the radius. The rate of inclination change is
${\mathrm{d}i \over \mathrm{d}t} = -\, 0.37 \pm 0.22$
$\mathrm{mas}\, \mathrm{yr}^{-1}$, a value similar to that
determined  from N-body simulations by Weinberg (2000), which
predicts the LMC disk precession and nutation 
to be due to  tidal torques generated by our Galaxy.

A third important conclusion is that the LMC disk has  a
considerable vertical thickness, in agreement also with the
numerical simulations of \citet{weinberg00}. The thickening of
the LMC disk is due to the gravitational interaction with the MW.
The ratio ${V\over\sigma} = 2.9 \pm 0.9$ is even lower than the
corresponding value for the MW thick disk (${V\over\sigma} \simeq
3.9$). The LMC disk is also flared. The best fit of the observed
velocity dispersion profile with isothermal disk models, whose
vertical density profile is proportional to
$\mathrm{sech}^{2}({z\over z_{0}})$, confirms the result found by
\citet{alves00} that the scale height must increase with radius.
The vertical thickening is also in agreement with the results
of \citet{olsen02}, who  argued that the LMC contains features
that extend up to 2.5 kpc out of the plane.

Let us note that recently some of these conclusions have been challenged by
\citet{nikolaev04}, whose analysis is based on a combination of the
results of the MACHO collaboration on the LMC Cepheides with the $2$MASS
All-Sky Release Catalog.

%
\subsection{LMC Dark Halo}
Another important conclusion of the van der Marel et al. analysis,
is connected with the LMC dark halo. Comparing the LMC total mass
enclosed in a sphere of $8.9$ kpc, dynamically inferred  to be
$(8.7 \pm 4.3) \times 10^9\, \mathrm{M}_{\sun}$, with respect to
the visible mass $M_{\mathrm{vis}}\approx 2.7 \times 10^9
\mathrm{M}_{\sun}$, and to the mass of neutral gas $\approx 0.5
\times 10^9 \mathrm{M}_{\sun}$ \citep{kim98}, \citet{marel02} have
concluded that the LMC must be surrounded by a dark halo with a
radius equal to the LMC tidal radius $r_{\mathrm{t}} = 15.0 \pm
4.5$ kpc, a value higher than that, $11\,\mathrm{kpc}$, estimated
by \citet{weinberg00}. Moreover, at a distance of 5 kpc from the
center of the LMC, the ratio between the tidal force and the LMC self
gravitational attraction is reduced by 20\%. Thus, one expects
that tidal effects influence deeply the structure of the LMC halo,
causing probably an elongation of it.

%
\section{Models} \label{sec:modelli}
We adopt the same coordinate systems and notations as in
\citet{marel02}. The position of any point in space is uniquely
determined by the two angles, right ascension $\alpha$ and
declination $\delta$,  and by its distance $D$. We introduce also
a cartesian coordinate system $(x,\, y,\, z)$ that has its origin
$O$ in the center of the LMC, whose coordinates are $(\alpha_{0}, \,
\delta_{0}, \, D_{0})$, with the $x$--axis antiparallel to the
right ascension axis, the $y$--axis parallel to the declination
axis, and the $z$--axis pointing towards the observer. To describe
the internal kinematics of the galaxy it is opportune to introduce
a second cartesian coordinate system $(x',\, y',\, z')$ that is
obtained from the system $(x,\, y,\, z)$ by a counterclockwise
rotation around the $z$--axis by an angle equal to the position
angle of the line of nodes $\Theta-{\frac{\pi}{2}}$, followed by a
rotation around the new $x'$--axis by an angle equal to the
inclination angle of the disk plane $i$. With this definition the
$(x',\, y')$ plane coincides with the equatorial plane of the LMC
galaxy, and the $x'$--axis is along the line of nodes.
%

\subsection{LMC Disk} \label{sec:disk}
Following the new results on the LMC morphology discussed in Sect.
\ref{sec:morphology}, we consider an elliptical isothermal flared disk
tipped by an angle $i = 34.7^{\circ} \pm 6.2^{\circ}$ with respect to the
sky plane, with the closest part in the north-east side. The
center of the disk coincides with the center of the bar located at
J2000 ($\alpha,\delta$) = ($5^h 27.6^m \pm 3.9^m, - 69.87^{\circ}
\pm 0.41^{\circ}$) and its distance from us is $D_{0} = 50.1 \pm
2.5 \, \mathrm{kpc}$. Here we take a bar mass
$M_{\mathrm{bar}}=1/5\,M_{\mathrm{disk}}$ \citep{sahu94,gyuk00},
with the visible mass in disk and bar
$M_{\mathrm{bar}}+M_{\mathrm{disk}}=(2.7\pm 0.6) \times 10^{9}\,
\mathrm{M}_{\sun}$ \citep{marel02}.

The vertical distribution of stars in an isothermal disk is
described by the $\mathrm{sech}^{2}$ function, therefore the
spatial density  of the disk is modelled by:
\begin{equation}
\rho_{\mathrm{d}}=\frac{{\cal N}\, M_{\mathrm{d}}}{4 \pi\, q\,
R_{\mathrm{d}}^{2}\, \zeta_{\mathrm{d}}(0)}\;
\mathrm{sech}^2\left( \frac{\zeta}{\zeta_{\mathrm{d}}(R)}\right)
 e^{-{\frac{1}{R_{\mathrm{d}}}}\sqrt{
\left(\frac{\xi}{q}\right)^2+
{\eta}^2}}~,
\end{equation}
where $q = 0.688$ is the ellipticity factor,
$R_{\mathrm{d}}=1.54\,\mathrm{kpc}$ is the scale length of the
exponential disk, $R$ is the radial distance from the center on
the disk plane. ${\zeta_{\mathrm{d}}(R)}$ is the \textit{flaring}
scale height, which rises from 0.27 kpc to 1.5 kpc at a distance
of 5.5 kpc from the center \citep{marel02},  and is given by
$${\zeta_{\mathrm{d}}(R)}=0.27+1.40
\,\tanh \left(\frac{R}{4}\right)~.
$$
${\cal N} = 0.2765$ is a normalization factor that takes into
account the flaring scale height.

The cartesian coordinates $(\xi,\, \eta,\, \zeta \equiv z')$ are
obtained from the system $(x',\, y',\, z')$ by  rotation around
the $z'$-axis by an angle equal to $(\Phi-\Theta -
{\frac{\pi}{2}})$, where $\Phi$ is the position angle of the LMC disk
major axis. In this way the $\xi - \eta$ plane coincides with the
equatorial disk plane and the $\eta$ axis is directed along  the
major axis of the elliptic disk.

The velocity dispersion is a crucial parameter for the
estimate of the microlensing rate. The kinematic studies of the LMC
disk have shown that measurements of the velocity dispersion along
the line of sight vary  between roughly $6$ and $30$ km/s,
according to the age of the tracers  and show little variation as
a function of the radius (see \citet{gyuk00} where a comprehensive
table is given). In particular,  younger populations
have a smaller velocity dispersion than older ones, as in the MW. In the picture
of \citet{marel02} within a distance of about 3 kpc from the
center of the LMC, the line of sight velocity dispersion (evaluated
for carbon stars) can be considered as constant with
$\sigma_{\mathrm{d,los}}=20.2\pm 0.5$ km/s. This represents our
choice for the  line of sight velocity dispersion of the disk
stars.

\subsection{LMC Bar}
\label{sec:bar}
An optical bar of roughly $3^{\circ} \times 1^{\circ}$
angular size stands out in the central region of the LMC. Unlike
the disk, the geometry of the bar region, in particular its
vertical structure, is not well constrained. Understanding how the
bar is displaced with respect to the disk is one of the last
information needed to complete the puzzle of the LMC morphology.

In Paper I we have described the bar by a gaussian density profile
following \citet{gyuk00}. In this paper we choose, instead, a bar
spatial density  that takes into account its boxy shape, as in
\citet{zhao00}:
\begin{equation}
\rho_{\mathrm{b}}=\frac{2\,M_{\mathrm{b}}}{\pi^{2} \,
R_{\mathrm{b}}^{2}\,\, \Xi_{\mathrm{b}}}\, e^{
-\left(\frac{\Xi}{\, \Xi_{\mathrm{b}}}\right)^{2}}
\,e^{-\,\frac{1}{R_{\mathrm{b}}^{4}}
\left(\Upsilon^{2}+\,\zeta^{2}\right)^{2}},
\end{equation}
where $\Xi_{\mathrm{b}}=1.2\,\mathrm{kpc}$ is the scale length of
the bar axis,  $R_{\mathrm{b}}=0.44\,\mathrm{kpc}$ is the scale
height along a circular section, and the cartesian coordinates
$(\Xi,\, \Upsilon,\, \zeta \equiv z')$ are obtained from the
system $(x',\, y',\, z')$ by  rotating around the $z'$-axis by an
angle equal to $(\Psi-\Theta)$, where $\Psi$ is the position angle
of the LMC bar major axis,  roughly $\simeq 120^{\circ}$
\citep{marel01b}. We fix it to $112.7^{\circ}$, so that the bar
major axis $\Xi$ coincides with the minor axis of the disk $\xi$.

\begin{figure}
 \resizebox{\hsize}{!}{\includegraphics{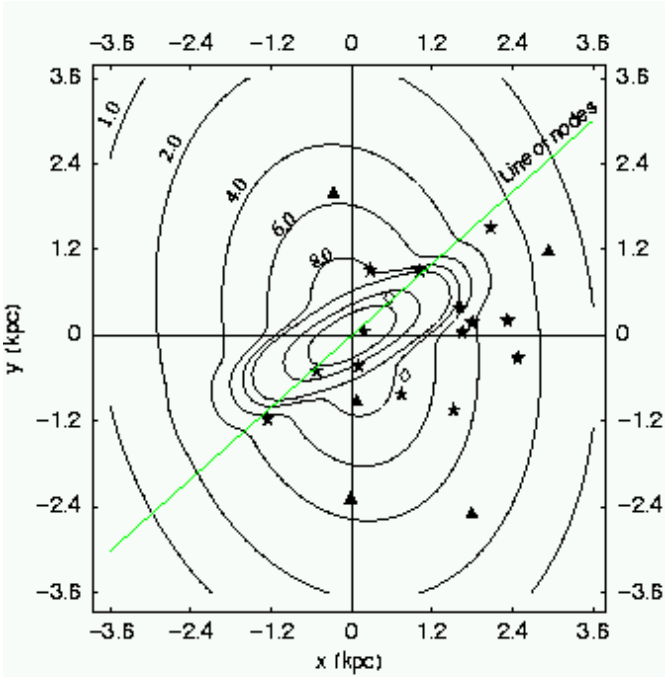}}
 \caption{Projection on the sky plane ($x-y$ plane) of the column density
 of the LMC disk and bar. The numerical values on the contours are in $10^{7}
 \mathrm{M}_{\sun}\mathrm{kpc}^{-2}$ units.
 The three innermost contours correspond to $10$, $20$ and $30\times 10^{7}
 \mathrm{M}_{\sun}\mathrm{kpc}^{-2}$. The locations of the MACHO
 (black stars and empty diamonds) and EROS
 (triangles) microlensing candidates are also shown.}
 \label{fig:column-density-plot}
\end{figure}

The column density, projected on the $x-y$ plane is plotted in
Fig. \ref{fig:column-density-plot}, giving a global view of the LMC
shape (disk and bar coplanar) for a terrestrial observer.
We indicate the direction of the line of nodes,
together with the positions of the microlensing events detected by
the MACHO (filled stars and empty diamonds) and EROS (filled
triangles)\footnote{The EROS group has published the detection of
5 microlensing events towards the LMC \citep{lasserre00}. The
event EROS2-LMC-5 is dubious as its light curve is asymmetric.
This candidate has actually been considered by MACHO as its
candidate LMC-26 and rejected as being a supernova
\citep{milsztajn03}.} collaborations. The maximum value of the column density,
$41.5\times 10^{7} \mathrm{M}_{\sun}\,\mathrm{kpc}^{-2}$, is
assumed in the center of the LMC. 
The positions of the MACHO microlensing events in this $x-y$ reference system are
reported in Table \ref{tab:gtecomp}.

In order to explore the consequences of different bar
geometries, besides from the coplanar configuration as in
\citet{gyuk00} and \citet{weinberg01}, we have considered two
other possible geometries. In particular, we drop the hypothesis
that the disk and bar components are dynamically connected. In a
first case, drawing inspiration from the paper of \citet{zhao00},
we rotate the bar around an axis through the center, orthogonal to
the plane defined by the bar axis and the line of sight. As a
second case we shift the bar towards the observer along the line
of sight as in \citet{nikolaev04}. In both cases we change
accordingly the scale parameters of the bar so as to keep its
observed projected shape on the sky plane fixed. For both
configurations we considered several values for the shift and the
rotation angle in order to study their influence on the
microlensing quantities. In the following, as an illustration, we
will give the results for two somewhat extreme cases, so that we
will get upper limits for the corresponding microlensing
quantities
\begin{itemize}
\item{} a translation towards the observer of $\sim 0.5$ kpc;
\item{} a rotation of $40^{\circ}$
with the south-east side towards the observer. In this case
the linearly growing  distance between bar and disk
is $\sim 1.0$ kpc at one scale length from the center.
\end{itemize}

For the velocity dispersion
of the stellar population of the bar, for which very
few data are available, we take into account the
qualitative results from numerical simulation that show
a general trend of a higher line of sight velocity dispersion in
the central region of the LMC \citep{zasov02}. We consider again
two somewhat extreme cases:
a line of sight velocity dispersion equal to that
of the disk, $\sigma_{\mathrm{b,los}}=20.2$ km/s, and
a second case with a higher value,
$\sigma_{\mathrm{b,los}}=30$ km/s.

\subsection{ LMC Halo}
We use two different models to describe the halo profile density:
a spherical halo (model S) and an ellipsoidal halo (model E). The
values of the parameters have been chosen so that the models have
roughly the same mass within the same radius.
\begin{itemize}
\item[$\bullet$] Model S.
In this model we neglect the tidal effects due to our Galaxy, and
we adopt  a classical pseudo-isothermal spherical  density
profile, which allows us to compare the results obtained in this
paper with the previous ones:
\begin{equation}
\rho_\mathrm{h,S}=\rho_{\mathrm{0,S}}
\left[1+\frac{R^{2}}{a^{2}}\right]^{-1} \theta(R_{\mathrm{t}}-R),
\end{equation}
where $a$ is the LMC halo core radius,  $\rho_{0,\mathrm{S}}$ the
central density, $R_{\mathrm{t}}$ a cutoff radius and $\theta$ the
Heaviside step function. We use $a=2$ kpc \citep{alcock00a}. We
fix the value for the mass of the halo within a radius of 8.9 kpc
equal to $5.5\times10^{9}\mathrm{M}_{\sun}$ \citep{marel02} that
implies $\rho_{0,\mathrm{S}}$ equal to
$1.76\times10^{7}\,\mathrm{M}_{\sun}\,\mathrm{kpc}^{-3}$. Assuming
a halo truncation radius, $R_{\mathrm{t}}= 15$ kpc
\citep{marel02}, the total mass of the halo is $\approx
1.08\times10^{10}\,\mathrm{M}_{\sun}$.

\item[$\bullet$] Model E.
In this model the MW tidal effects induce a oblate spheroidal
shape to the halo. We describe the halo density profile as
\begin{equation}
\rho_\mathrm{h,E}=\rho_\mathrm{0,E}\left( 1+\frac{1}{a^2} \left(
\frac{\xi^2}{q_{1}^{2}}+\eta^2+\zeta^2
\right) \right)^{-1},
\end{equation}
with  $a=2$ kpc, and the truncation radius  $R_{\mathrm{t}}=15$
kpc. For a LMC dark mass equal to
$5.5\times10^{9}\,\mathrm{M}_{\sun}$ within a radius of 8.9 kpc,
the central density is
$2.20\times10^{7}\,\mathrm{M}_{\sun}\,\mathrm{kpc}^{-3}$, so that
the LMC dark mass within the truncation radius turns out to be
$\approx 1.06\times10^{10}\,\mathrm{M}_{\sun}$.
\end{itemize}
%

\subsection{Galactic halo}

We will consider for the galactic halo a spherical model with density
profile given by:
\begin{equation}
\rho_\mathrm{GH}=\rho_{\mathrm{0}}
\frac{R_{\mathrm{C}}^{2} + R_{\mathrm{S}}^{2}}{R_{\mathrm{C}}^{2}+R^{2}},
\end{equation}
where $R$ is the distance from the galactic center,
$R_{\mathrm{C}} = 5.6 \,\,\mathrm{kpc}$ \citep{primack} is the
core radius, $R_{\mathrm{S}} = 8.5 \, \,\mathrm{kpc}$ is the
distance of the sun from the galactic center and
$\rho_{\mathrm{0}} = 7.9\times 10^{6}\,
\mathrm{M}_{\sun}\,\mathrm{kpc}^{-3}$ is the mass density in the
solar neighbourhood.

\section{Calculation of microlensing parameters} \label{sec:mlpar}

As outlined in the introduction, here and in the following
sections we want to fully exploit all the consequences of the
geometry of the LMC,  with the aim to shed
new light on the still puzzling nature and location of the lenses
detected in the microlensing surveys. In particular we study here
two microlensing quantities, the optical depth and the
microlensing rate.

In our analysis, whenever we need to compare models and predictions
with observational results, we are going to use those presented by the
MACHO collaboration only. The main reason is that this team has
provided a complete description of their microlensing detection
efficiency.

From the total set of 17 MACHO microlensing
events\footnote{The MACHO team actually claims the detection of
13--17 events according to the chosen selection criteria. In the
following we shall consider for our purposes the largest set.} 
one must exclude the event LMC--22, that is very likely a
supernova of long duration or an active galactic nucleus in a
galaxy at red-shift $z=0.23$ \citep{alcock01c}.

We have detailed information on LMC--5: it is due to a lens
located in the Galactic disk. Indeed the lens proper motion has
been observed with the HST \citep{alcock01a} and the lens mass
determined to be either $\simeq 0.04$ M$_{\sun}$ or in the range
0.095 - 0.13 M$_{\sun}$, so that it is a true brown dwarf or a
M4-5V spectral type low mass star. The other stars which have been
microlensed were also observed by the HST, but no other lenses
have been detected. This result is also confirmed by the analysis
made by \citet{vonhippel03}, by using optical and near infrared
photometry on a subset of five lensing sources which are LMC main
sequence stars or slightly evolved subgiants, (LMC--4, LMC--6,
LMC--8, LMC--9, LMC--14). Their analysis rules out for main
sequence lens masses $\ge 0.1$ M$_{\sun}$ for distances out to 4
kpc.

The events LMC--9 and LMC--14 are known to be due to lenses
belonging to the LMC itself, i.e. to the bar, disk or halo component.
The latter event has been recognized thanks to the double source
\citep{alcock01b}, while the first shows the characteristic
caustic crossing signature of a double lens \citep{alcock00b}. For
this reason we exclude the event LMC--9 from the following
analysis, because we are interested in the study of an homogeneous
set of single lens events.

We remark that LMC--5, LMC--9 and LMC--14 are the only events for which it has
been possible to make a determination of the location of the lens.
In Fig. \ref{fig:column-density-plot} the events LMC--5 and LMC--9 are
indicated with a special symbol, an empty diamond, whereas LMC--22
is not reported in the plot.

\begin{figure}
 \resizebox{\hsize}{!}{\includegraphics{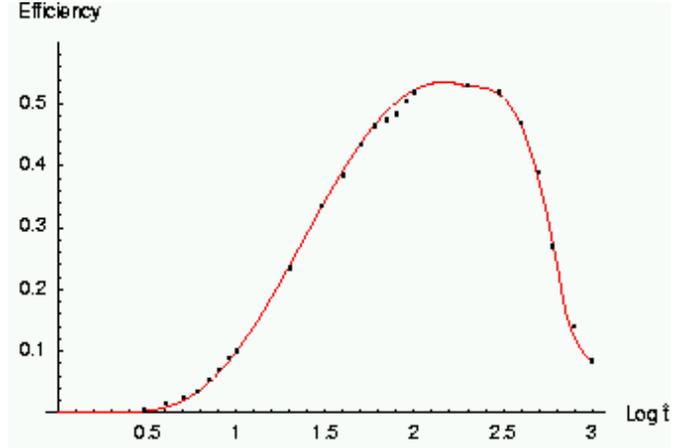}}
 \caption{Plot of the interpolating function  ${\cal E}({\hat t})$ of the
 MACHO detection efficiency.}
 \label{fig:efficiency}
\end{figure}

We use the  interpolating function ${\cal E}({\hat t})$ of the
microlensing detection efficiency, calculated by the MACHO
collaboration, as a function of event timescale ${\hat t}$ in the
interval $1 - 1000$ days (Fig. 5 of \citet{alcock00a}). It is
shown  in Fig. \ref{fig:efficiency} by the continuous line, together
with some points of the MACHO efficiency, for comparison. Let us
recall that the MACHO definition of the duration time $\hat t$ is
twice the Einstein time $T_{\mathrm{E}}$, the parameter we use in
this paper. We get the following expression for ${\cal E}({\hat
t})$:
\begin{eqnarray}\label{efficiencyEq}
&{\cal E}({\hat t}) = &f_{1}({\hat
t})\,\theta[2.3-{\mathrm{Log}}\,{\hat t}]\,
\theta[{\mathrm{Log}}\,{\hat t}] +\nonumber \\
&&f_{2}({\hat t})\,\theta[{\mathrm{Log}}\,{\hat t}-2.3]\,
\theta[2.85-{\mathrm{Log}}\,{\hat t}] + \nonumber  \\
&&f_{3}({\hat t})\,\theta[{\mathrm{Log}}\,{\hat t}-2.85]\,
\theta[3-{\mathrm{Log}}\,{\hat t}] \; ,
\end{eqnarray}
with $\theta$ the Heaviside function, and
\begin{eqnarray}
&f_{1}({\hat t}) = &{\frac{{\mathrm{Log}}^{3}\,{\hat t}}{{\hat
t}^{1.97}}}\left[8\, {\mathrm{Log}}^{5}\,{\hat t}+{\frac{49}{40}}
{\mathrm{Log}}^{8}\,{\hat t}\right]\\
&f_{2}({\hat t})=&{\frac{f_{1}(10^{2.3})-{\frac{125}{64}}\left(
{\mathrm{Log}}\,{\hat t}-2.3\right)^{3}}{10^{\left(2.3-
{\mathrm{Log}}\,{\hat t}\right)^{4}}}}  \\
&f_{3}({\hat t})=&0.06+\left[f_{2}(10^{2.85})-0.06\right]
e^{10(2.85 -{\mathrm{Log}}\,{\hat t})} \; .
\end{eqnarray}

As described in the previous section, the LMC has its own halo, with a
tidal radius $R_{\mathrm{t}} = 15.0 \pm 4.5$ kpc. Assuming that
the galactic halo extends to at least 65 kpc, a fraction of the
LMC halo total mass, enclosed in a sphere of radius
$R_{\mathrm{t}}$, is attributable to the galactic halo. A simple
calculation gives the galactic halo mass enclosed in a sphere
centered on the LMC with a radius of 8.9 kpc to be $9.7\times
10^{8}\,\mathrm{M}_{\sun}$, a value corresponding to a sensible
fraction of the LMC halo mass, of the order of $\simeq 17\%$. We
will take into account this fact, by properly correcting the value
of the central density of the LMC halo: for the spherical model
the value is decreased to $\rho_{0,\mathrm{S}} = 1.45 \times
10^{7}\, \mathrm{M}_{\sun}\,\mathrm{kpc}^{-3}$ and the
corresponding total LMC halo mass inside the tidal radius
$R_{\mathrm{t}}$ is $\approx 8.86\times10^{9}\,\mathrm{M}_{\sun}$;
for the ellipsoidal model the value is decreased to
$\rho_{0,\mathrm{E}} = 1.81 \times 10^{7}\,
\mathrm{M}_{\sun}\,\mathrm{kpc}^{-3}$ and the corresponding total
LMC halo mass inside the tidal radius $R_{\mathrm{t}}$ is $\approx
8.7\times10^{9}\,\mathrm{M}_{\sun}$.

%
\subsection{Optical depth} \label{sec:tau}
We now discuss the results obtained for the optical depth. The
computation is made by weighting the optical depth with respect to
the  distribution of the source stars along the line of view. We
use Eq. (7), Sect. 3.1, of Paper I, that we report here for self
consistency reasons:
\begin{equation}\label{weightedOD}
\tau = {\frac{4\pi G}{c^{2}}}
{\frac{\int_{0}^{\infty}\left[\int_{0}^{D_{\mathrm{os}}}
{\frac{D_{\mathrm{ol}}(D_{\mathrm{os}}-D_{\mathrm{ol}})}
{D_{\mathrm{os}}}}\rho_{\mathrm{l}}
\,dD_{\mathrm{ol}}\right]\rho_{\mathrm{s}}\,
dD_{\mathrm{os}}}{\int_{0}^{\infty}\rho_{\mathrm{s}}\,
dD_{\mathrm{os}}}}\;,
\end{equation}
where $\rho_{\mathrm{l}}$ denotes the mass density of the lenses,
$\rho_{\mathrm{s}}$ the mass density of the sources,
$D_{\mathrm{ol}}$ and $D_{\mathrm{os}}$, respectively, the
distance observer-lens and observer-source.

\begin{figure}
 \resizebox{\hsize}{!}{\includegraphics{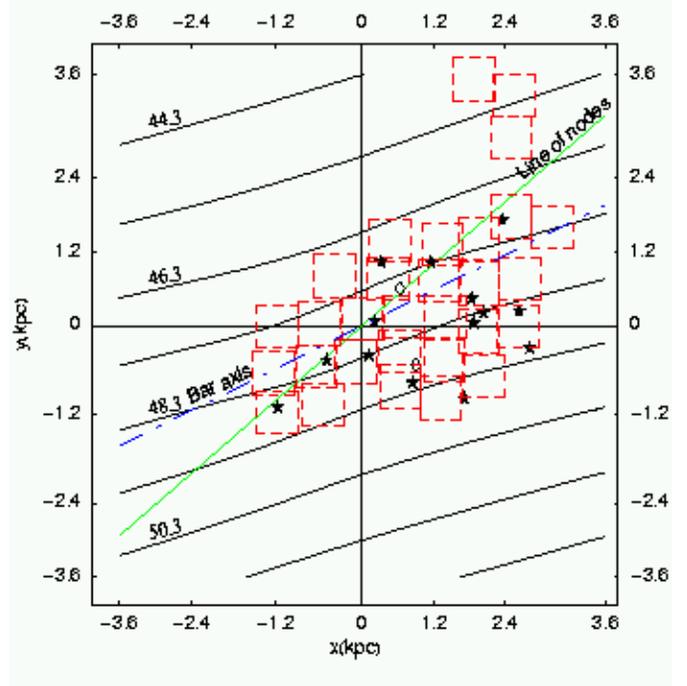}}
 \caption{Contour map of the optical depth for
 lenses in the galactic halo. The locations of the
 MACHO fields and of the microlensing candidates are also shown.
 The numerical values are in $10^{-8}$ units.}
 \label{fig:GH}
\end{figure}

\begin{figure}
 \resizebox{\hsize}{!}{\includegraphics{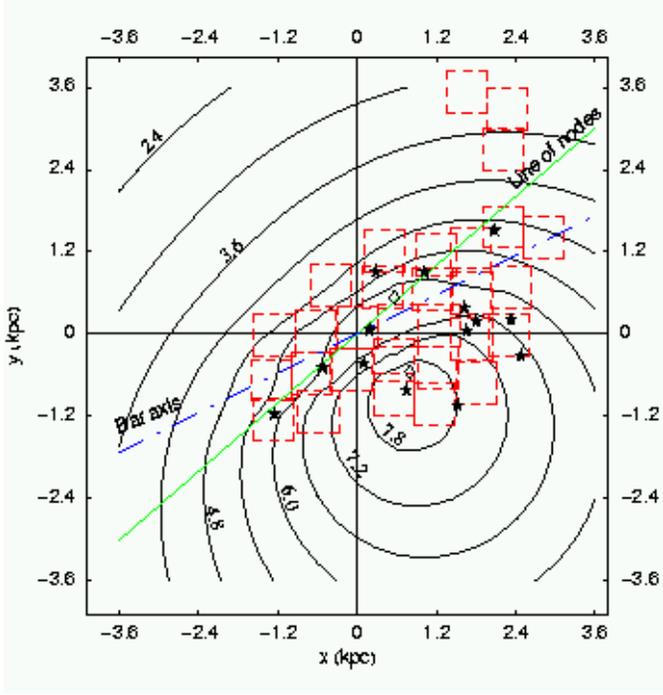}}
 \caption{Model S: contour map of the optical depth for
 lenses in the LMC halo. The locations of the
 MACHO fields and of the microlensing candidates are also shown.
 The numerical values are in $10^{-8}$ units.}
 \label{fig:LMCHS}
\end{figure}

\begin{figure}
 \resizebox{\hsize}{!}{\includegraphics{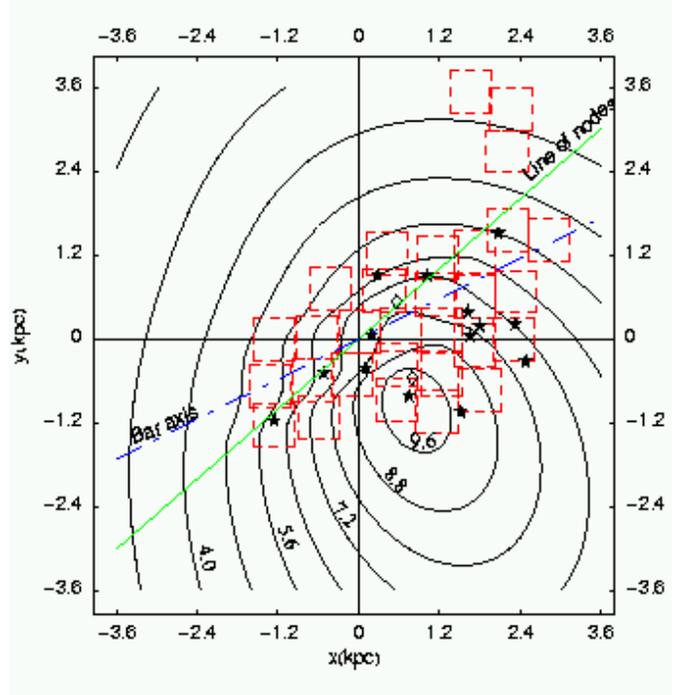}}
 \caption{Model E: contour map of the optical depth for
 lenses in the LMC halo. The locations of the
 MACHO fields and of the microlensing candidates are also shown.
 The numerical values are in $10^{-8}$ units.}
 \label{fig:LMCHE}
\end{figure}

\begin{figure}
\resizebox{\hsize}{!}{\includegraphics{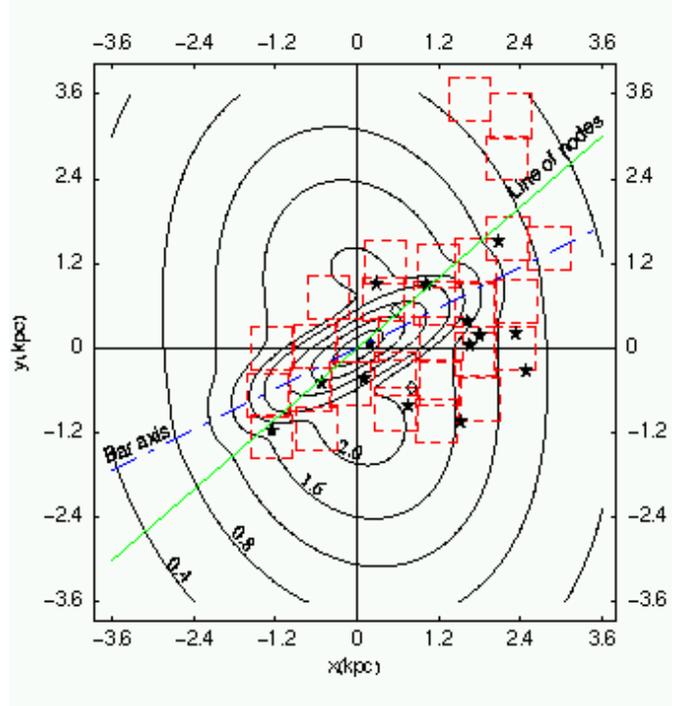}}
\caption{Contour map of the optical depth for self lensing.  The
 locations of the MACHO fields and of the microlensing candidates
 are also shown. The numerical values are in $10^{-8}$ units.
 The innermost contours correspond to
 values $2.4\times 10^{-8}$, $3.2\times 10^{-8}$, $4.0\times 10^{-8}$
 and $4.6\times 10^{-8}$ respectively.}
\label{fig:SL}
\end{figure}

The contour maps reported in Figs.  \ref{fig:GH}, \ref{fig:LMCHS},
\ref{fig:LMCHE} and  \ref{fig:SL} refer to the case of coplanarity between
bar and disk. In Fig. \ref{fig:GH} is reported the optical depth
contour map for lenses in the Galactic halo, in the hypothesis
that all the Galactic dark halo consists of compact lenses,
together with the positions of the MACHO fields (see Fig.
\ref{fig:asymmetry} for the numeration of the fields), the
microlensing events and the van der Marel line of nodes. We
observe that almost all the fields (except three of them) fall
between the contour lines corresponding to $\tau = 46.3 \times
10^{-8}$ and $\tau = 49.3 \times 10^{-8}$. As expected the optical
depth due to the galactic halo is a slowly variable function, and
presents a slight near-far asymmetry: moving from the nearer to
the farther fields along a line passing through the center and
perpendicular to the line of nodes, the increase of the optical
depth is of the order of $\approx 6\%$.

In Figs. \ref{fig:LMCHS} and  \ref{fig:LMCHE} we report the optical depth
contour maps for lenses belonging to the halo of the LMC,
assuming a spherical and an ellipsoidal shape, respectively,
in the hypothesis that all the LMC dark halo consists of compact
lenses. A striking feature of both maps is the strong near-far
asymmetry.

For the Model S, the maximum value of the optical depth,
$\tau_{\mathrm{max,S}} \simeq 8.05 \times 10^{-8}$, is assumed in
a point falling in the field number 13, belonging to the fourth
quadrant, at a distance of $\simeq 1.27$ kpc from the center. The
value in the point symmetrical with respect to the center,
belonging to the second quadrant and falling about at the upward
left corner of the field 82, is $\tau_{\mathrm{S}} \simeq 4.30
\times 10^{-8}$. The increment of the optical depth is of the
order of $\approx 87\%$,  moving from the nearer to the farther
fields.

The same happens for the model E: the maximum value of the optical
depth, $\tau_{\mathrm{max,E}} \simeq 9.88 \times 10^{-8}$,  higher
than the previous one, is found at about the same point, at the
same distance from the center. In the symmetrical point with
respect to the center, belonging to the field 82, the optical
depth is $\tau_{\mathrm{E}} \simeq 5.05 \times 10^{-8}$. The
ellipsoidal shape of the LMC halo gives rise to a further
enhancement of the near-far asymmetry, with an increase of the
optical depth by $\approx 95\%$.

One can draw advantage from the different asymmetric behaviour of
the optical depth in the two cases of lenses in the galactic halo
or in the LMC halo, both to confirm the existence of a proper LMC
halo and to disentangle the microlensing events due to the
Galactic halo from the ones due to the LMC halo. To this end, a
good observation strategy would help, the goal being to allow the
analysis of asymmetry of microlensing events belonging to two
equivalent regions, placed symmetrically  with respect to the line
of nodes. An example of the kind of analysis and of results one
could obtain is given in Sect. \ref{sec:asimmetria}.

In Fig. \ref{fig:SL} we report the optical depth contour map for self
lensing, i. e. for events where both the sources and the lenses
belong to the stellar population of the disk and/or the bar of the LMC. 
As expected, there is almost\footnote{We note the peculiar ``ear'' shape of the
contour line for $\tau = 2 \times 10^{-8}$ due to the disk
flaring.} no near-far asymmetry and the maximum value of the
optical depth, $\tau_{\mathrm{max}} \simeq 4.90 \times 10^{-8}$,
is reached in the center of the LMC. The optical depth then  rapidly
decreases, when moving, for instance, along a line going through
the center and perpendicular to the minor axis of the elliptical
disk, that coincides also with the major axis of the bar. In a
range of about only $0.80 \,\mathrm{kpc}$ the optical depth
quickly falls to $\tau \simeq 2 \times 10^{-8}$, and afterwards it
decreases slowly to lower values.

The calculated value of the optical depth in the center  seems at
first glance to be  in contradiction with the value one gets using
the Gould inequality \citep{gould95}:
\begin{equation}
\tau \le 2{\frac{<v^{2}>}{c^{2}}} \sec^{2}i\approx 1.34 \times
10^{-8}.
\end{equation}
In fact, this inequality is derived making some simplifying
assumptions which do not fully apply to our choice  of the density
profiles for the disk and the bar. In Sect. \ref{sec:gould} we shall derive an
improved version of the Gould inequality, which will also apply
to our density profile and leads to $\tau \le 6 \times 10^{-8}$,
which does not contradict our above estimated value.

Furthermore, we recall that the Gould inequality is obtained
under the hypothesis that the LMC is a virialized system,
that quite likely is note the case for some of its components.
This should of course be taken into account
when using it in comparison with the observations.

As expected, the changes of the geometry as discussed in
Sect. \ref{sec:bar}, where we allow for a non coplanar morphology of the bar
with respect to the disk, enhance the self--lensing optical depth 
in the bar region considerably (up to $\sim$ 50\%).  On the other
hand, the changes for the optical depth for lenses in the MW halo
are negligible (at maximum $\sim$ 1\%). For lenses in the LMC halo
the variations, in the innermost region of the LMC, can be rather
large (up to $\sim$ 20\%). However, the main feature that interests
us, the near--far asymmetry due to the disk inclination, is not
altered. Besides, as we are here mainly concerned with the
self--lensing case, we do not discuss them any further. In any case,
outside the bar region, these differences drop abruptly to zero.

\begin{figure}
 \resizebox{\hsize}{!}{\includegraphics{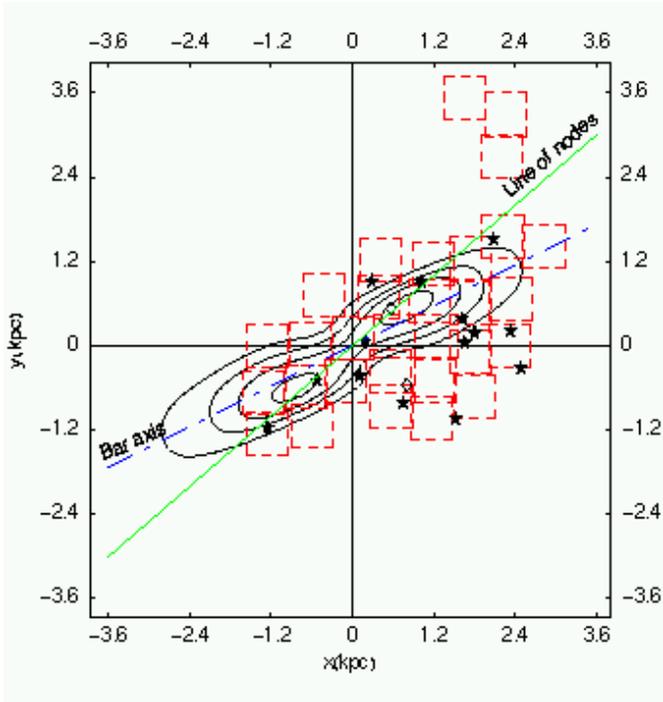}}
 \caption{Contour map of the difference between the self--lensing
  optical depth calculated in the case of $+ 40^{\circ}$ rotated bar
  and the one with coplanar bar and disk. Going from the
  outer to the inner part, the contours correspond to
   increasing values of the  difference: $0.1, \, 0.6, \, 1.1 \,
   {\mathrm{and}}\, 1.6 \times 10^{-8}$. The difference is null
   outside the bar region.}
 \label{fig:SLBarRotated}
\end{figure}

For a rotated bar, Fig. \ref{fig:SLBarRotated}, the  increment of
the self--lensing optical depth, with respect to the symmetric coplanar case, 
is slightly asymmetric with respect to the bar major axis.
This is due to the different variations of the source--lens distances
between the west side, where the sources are in the bar and the
lenses in the disk, and the east side, where  the opposite
happens. The relative increment can be as large as $\approx  50 \, \%$.

\begin{figure}
 \resizebox{\hsize}{!}{\includegraphics{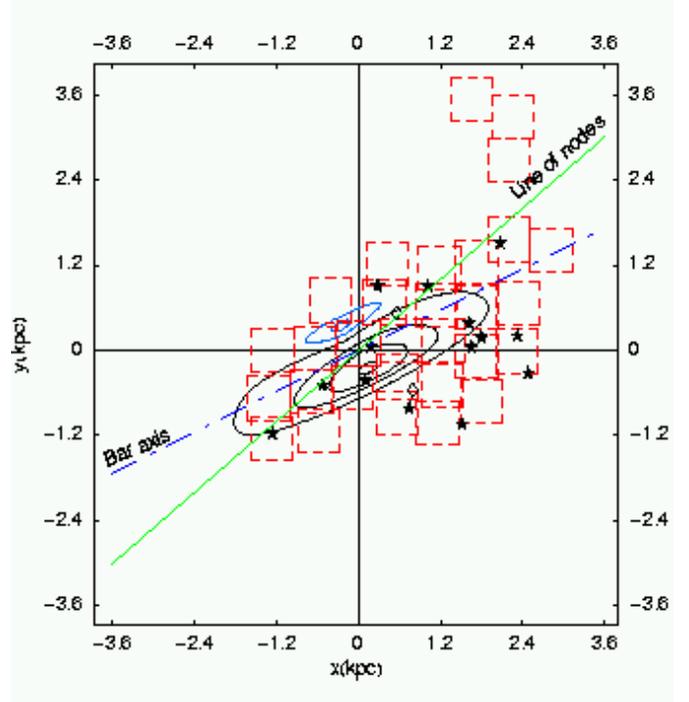}}
 \caption{Contour map of the difference between the self--lensing
  optical depth calculated in the case of $0.5$ kpc shifted bar
  and the one with coplanar bar and disk. Going from the
  outer to the inner part, the black contours correspond to
   increasing positive values of the  difference: $+0.1, \, +0.5, \, +0.9, \,
   {\mathrm{and}}\, +1.3 \times 10^{-8}$; the lighter contours to negative
   values of the difference: $-0.1, \,
   {\mathrm{and}}\, -0.2 \times 10^{-8}$.}
 \label{fig:SLbarShifted}
\end{figure}

When we allow for a translation of the bar, Fig. \ref{fig:SLbarShifted}, 
the variation of the optical depth,
with respect to the coplanar case, is
higher in the region below the bar major axis (south-west side),
even if the projected density of the bar is the same as shown in
Fig. \ref{fig:column-density-plot}. This is induced by the inclination
of the disk plane, that gives rise to increasing distances between
the bar and the disk plane for lines of sight  towards the lower
part of the bar. In this case we find also a small
region with a negative difference of the optical depth, shown by
the lighter contour lines (north-east side). The relative increment
can be as large as $\approx 35 \, \%$. 

It is interesting to note that a larger statistics of observed
events in the central region might eventually allow one to
discriminate between the different bar models.

We notice also that, due to the loss of the symmetry, the Gould
inequality no longer applies in these cases.

%
\subsection{Self lensing microlensing event rate}
The distribution
$\frac{\mathrm{d}\Gamma}{\mathrm{d}\,T_{\mathrm{E}}}$, the
differential rate of microlensing events with respect to the
Einstein time $T_{\mathrm{E}}$, allows one to estimate the expected
typical duration and the expected number of the microlensing
events. In this paper we evaluate the microlensing rate in the
self--lensing configuration, i. e. lenses and sources both in the
disk and/or in the bar of the LMC. We take into account also the
transverse motion of the Sun and of the source stars.

We assume that, to an observer comoving with the LMC center, the
velocity distribution of the source stars and lenses have a
Maxwellian profile\footnote{The Maxwellian profile of the
velocity distribution is the first term of a series expansion in
terms of Gauss--Hermite moments \citep{marel93,gerhard93}. See
Section $3.2$ of Paper I.}. Moreover, since the component parallel
to the line of view is irrelevant to microlensing, we  integrate
over the parallel component. Therefore, the two--dimensional
transverse velocity distribution of the source and lens stars is:
\begin{equation}\label{starDistribution}
f({\tilde{\vec{v}}}_{\perp})\,\mathrm{d}^{2}{\tilde v}_{\perp}=
{\frac{1}{2\,\pi\sigma_{\mathrm{los}}^{2}}}\,
e^{\left({-{\frac{{\tilde{{v}}}_{\perp}^{2}}{2\,\sigma_{\mathrm{los}}^{2}}}}\right)}\,
{\tilde v}_{\perp}\,\mathrm{d}{\tilde
v}_{\perp}\,\mathrm{d}\varphi \; ,
\end{equation}
where we have used cylindrical coordinates\footnote{Eq. \ref{starDistribution}
gives the correct  mean square transverse
velocity ${<\tilde{{v}}}_{\perp}^{2}>=2 \sigma_{\mathrm{los}}^{2}$.}
d$^{2}{\tilde
v}_{\perp}= {\tilde v}_{\perp}\,\mathrm{d}{\tilde
v}_{\perp}\,\mathrm{d}\phi$.

With respect to an observer comoving with the LMC center, the
transverse velocity of the microlensing tube at position
$D_{\mathrm{ol}}=x\,D_{\mathrm{os}}$ ($0\le x \le 1$) is given by
${\tilde{\vec{v}}}_{\mathrm{t}\perp}=
(1-x)\,{{\tilde{\vec{v}}}_{\sun\perp}} + x\,
{\tilde{\vec{v}}}_{\mathrm{s}\perp}$, where
$\tilde{\vec{v}}_{\mathrm{s}\perp}$ is the velocity of the source,
and $\tilde{\vec{v}}_{\sun\perp}$ represents the Sun velocity in
the plane orthogonal to the line of sight, as measured by an
observer comoving with the LMC center. It results
$|{\tilde{\vec{v}}}_{\sun\perp}|\simeq 406$ km/s \citep{marel02}.

Let us consider a segment of cylindrical ring at position $x$, of
length d$D_{\mathrm{ol}}$, with radius equal to the Einstein radius
$$R_{\mathrm{E}}= \left({\frac{4\, G\, \mathrm{M}_{\sun}}{c^{2}}}
\,{\mu\, D_{\mathrm{os}}\, x(1-x)}\right)^{\frac{1}{2}} \; ,
$$
angular amplitude d$\alpha$ and thickness $({\hat{\vec
v}}_{\mathrm{l}\perp}\cdot\hat n)\, \mathrm{d}t={\hat
v}_{\mathrm{l}\perp}\,\cos\theta\, \mathrm{d}t$, where
$-\,{\frac{\pi}{2}}\le\theta\le{\frac{\pi}{2}}$ is the angle
between the inner normal to the cylindrical ring segment and the
vector ${\hat{\vec v}}_{\mathrm{l}\perp}$. ${\hat{\vec v}}_{\mathrm{l}\perp}$ 
represents the lens velocity in the plane orthogonal to the line of sight, as measured
by an observer comoving with the microlensing tube at position
$x$. We use solar mass units, defining $\mu =
{\frac{M}{\mathrm{M}_{\sun}}}$, where $M$ is the lens mass.  The
number of lenses with mass in the range $(\mu,\,
\mu+\mathrm{d}\mu)$ inside the volume element of the cylindrical
ring segment is equal to the product of the differential number
density of the lenses ${\frac{\mathrm{d}\,
n_{\mathrm{l}}}{\mathrm{d}\, \mu}}\,\mathrm{d}\, \mu$ by the
volume element ${\hat{ v}}_{\mathrm{l}\perp}\cos\theta\,
R_{\mathrm{E}}\, \mathrm{d}t\, \mathrm{d}\alpha\,
\mathrm{d}D_{\mathrm{ol}}$. The fraction entering in the time
interval d$t$ in the microlensing tube with transverse velocity
${\hat{\vec v}}_{\mathrm{l}\perp}$, is given by those lenses that
with respect to the LMC reference frame have velocity
\begin{equation}
{\tilde{\vec v}}_{\mathrm{l}\perp}={\hat{\vec
v}}_{\mathrm{l}\perp}+\tilde{\vec{v}}_{\mathrm{t}\perp}
={\hat{\vec v}}_{\mathrm{l}\perp}+
(1-x)\,{\tilde{\vec{v}}_{\sun\perp}}+x\,
\tilde{\vec{v}}_{\mathrm{s}\perp} \; .
\end{equation}
In the self--lensing case, $x\simeq 1$, and to first approximation
we can neglect the term proportional to $(1-x)$. As a consequence
the velocity distribution with respect to the microlensing tube,
in the general case of two different values for the velocity
dispersion, is given by:
\begin{eqnarray}
{\frac{{\hat{v}}_{\mathrm{l}\perp}
\,\mathrm{d}{\hat{v}}_{\mathrm{l}\perp}\,\mathrm{d}\theta }{2\, \pi\,
\left[\rho_{\mathrm{d}}(x)+\rho_{\mathrm{b}}(x)
\right]}}\left[e^{\left(-{
\frac{{\hat{v}}_{\mathrm{l}\perp}^{2}+2\, x\,
{\hat{v}}_{\mathrm{l}\perp}\, {\tilde{v}}_{\mathrm{s}\perp}\,
\cos(\alpha-\theta) + x^{2}\,
{\tilde{v}}_{\mathrm{s}\perp}^{2}}{2\,
\sigma_{\mathrm{d}}^{2}}}\right)}\right.\cdot &  \nonumber\\
\left. {\frac{\rho_{\mathrm{d}}(x)}{
\sigma_{\mathrm{d}}^{2}}}+ e^{\left(-{
\frac{{\hat{v}}_{\mathrm{l}\perp}^{2}+2\, x\,
{\hat{v}}_{\mathrm{l}\perp}\, {\tilde{v}}_{\mathrm{s}\perp}\,
\cos(\alpha-\theta) + x^{2}\,
{\tilde{v}}_{\mathrm{s}\perp}^{2}}{2\,
\sigma_{\mathrm{b}}^{2}}}\right)}{\frac{\rho_{\mathrm{b}}(x)}{
\sigma_{\mathrm{b}}^{2}}}\right]&= \nonumber\\
{\frac{{\hat{v}}_{\mathrm{l}\perp}
\,\mathrm{d}{\hat{v}}_{\mathrm{l}\perp}\,\mathrm{d}\theta }{2\, \pi\,
\left[\rho_{\mathrm{d}}(x)+\rho_{\mathrm{b}}(x) \right]}}\,
h(\rho_{\mathrm{d}},\sigma_{\mathrm{d}};\,
\rho_{\mathrm{b}},\sigma_{\mathrm{b}};\,
x, {\hat{v}}_{\mathrm{l}\perp}, {\tilde{v}}_{\mathrm{s}\perp};\,
\alpha,\theta)\, &
\end{eqnarray}
where $\sigma_{\mathrm{d}}\equiv\sigma_{\mathrm{d, los}}$,
$\sigma_{\mathrm{b}}\equiv\sigma_{\mathrm{b, los}}$ and
$\alpha$ is the angle that the segment of cylindrical ring
at position $x$, of angular amplitude d$\alpha$, forms with the
direction of the orthogonal component to the line of sight of the
velocity vector of the source star,
$\tilde{\vec{v}}_{\mathrm{s}\perp}$.

We need now to specify the form of the number density. Assuming
that the mass distribution of the lenses is independent of their
position in the LMC ({\it factorization hypothesis}
\citep{derujula95}), the lens number density per unit mass is
given by
\begin{equation}
{\frac{\mathrm{d}n_{\mathrm{l}}}{\mathrm{d}
\mu}}={\frac{\rho_{\mathrm{d}}+\rho_{\mathrm{b}}}{\mathrm{M}_{\sun}}}\,
{\frac{\mathrm{d} n_{0}}{\mathrm{d} \mu}},
\end{equation}
where we use ${\frac{\mathrm{d} n_{0}}{\mathrm{d} \mu}}$  as given
in \citet{chabrier01}. We consider both the power law and the
exponential initial mass functions\footnote{We have used the same
normalization as in Paper I with the mass varying in the range
0.08 to 10 M$_{\sun}$.}. However, we find that our results do not
depend strongly on that choice and hereafter, we will discuss the
results we obtain by using the exponential IMF only.

Let us note that, considering the experimental conditions for the
observations of the MACHO team, we use as  range of variability
for the lens masses  $0.08 \le\mu \le 1.5$. Namely the lower limit
is fixed by the requirement that the lenses belong to the
luminous component of the LMC, while the upper limit is fixed by the
requirement that the lenses are not resolved stars\footnote{We
have checked that the results are insensitive to the precise upper
limit value. This is also confirmed by the following analysis on
the expected value for the mass of the lenses.}.

The total differential rate d$\Gamma$ at which lenses  enter the portion
of the microlensing tube at position $x$, along a fixed line of
sight, is given by \citep{derujula,griest}:
\begin{eqnarray}
\mathrm{d}\,\Gamma &=&{\frac{{\hat v}_{\mathrm{l}\perp}^{2}\,
d{\hat v}_{\mathrm{l}\perp}}{4\, \pi^{2} {\cal N}
\,\mathrm{M}_{\sun} }}\,\int_{0}^{2\, \pi}\mathrm{d}\varphi
\int_{-{\frac{\pi}{2}}}^{{\frac{\pi}{2}}}\, \mathrm{d} \theta \,
\,
{{\cos\theta}}\, \cdot \nonumber\\
&&
\int_{d_{\mathrm{min}}}^{d_{\mathrm{max}}}
D_{\mathrm{os}}
\,dD_{\mathrm{os}}
\int_{\frac{d_{\mathrm{min}}}{D_{\mathrm{os}}}}^{1}\mathrm{d}x
\,  \int_{0.08}^{1.5} \mathrm{d} \mu\, {\frac{\mathrm{d}
n_{0}}{\mathrm{d} \mu}}\,
R_{\mathrm{E}}\,
\cdot \nonumber\\
&&\int_{0}^{\infty}\mathrm{d}{\tilde{v}}_{\mathrm{s}\perp}
{{\tilde{v}}_{\mathrm{s}\perp}}
\left[{\frac{\rho_{\mathrm{d}}(D_{\mathrm{os}})}{
\sigma_{\mathrm{d}}^{2}}}\,e^{-{
\frac{{\tilde{v}}_{\mathrm{s}\perp}^{2}}{2\,
\sigma_{\mathrm{d}}^{2}}}}+{\frac{\rho_{\mathrm{b}}(D_{\mathrm{os}})}{
\sigma_{\mathrm{b}}^{2}}}\,e^{-{
\frac{{\tilde{v}}_{\mathrm{s}\perp}^{2}}{2\,
\sigma_{\mathrm{b}}^{2}}}}\right]
\cdot \nonumber\\
&&\int_{0}^{2\, \pi} \mathrm{d}\alpha\,
h(\rho_{\mathrm{d}},\sigma_{\mathrm{d}};\,
\rho_{\mathrm{b}},\sigma_{\mathrm{b}};\,
x, {\hat{v}}_{\mathrm{l}\perp}, {\tilde{v}}_{\mathrm{s}\perp};\,
\alpha,\theta)
\end{eqnarray}
where
$$
{\cal N} = \int_{d_{\mathrm{min}}}^{d_{\mathrm{max}}}
\mathrm{d}D_{os}\,\left[\rho_{\mathrm{d}}(D_{\mathrm{os}})
+\rho_{\mathrm{b}}(D_{\mathrm{os}})\right],
$$
having assumed that the number of detectable stars varies with the
distance as $D_{\mathrm{os}}^{-2}$. The integration limits
$d_{\mathrm{min}}$, $d_{\mathrm{max}}$ represent the distances
from the observer of the intersection points of the line of sight
with the LMC tidal surface.

Finally, as we are interested in the distribution
$\frac{\mathrm{d} \Gamma}{\mathrm{d} T_{\mathrm{E}}}$, we change
variable from ${\hat v}_{\mathrm{l}\perp}$ to $T_{\mathrm{E}}$,
bearing in mind that ${\hat
v}_{\mathrm{l}\perp}={\frac{R_{\mathrm{E}}}{T_{\mathrm{E}}}}$.
After integration on $\varphi$, $\theta$ and
$\alpha$, and taking into account the detection efficiency function, Eq.
(\ref{efficiencyEq}), we obtain:
\begin{eqnarray}
\left({\frac{\mathrm{d}\Gamma}{\mathrm{d}T_{\mathrm{E}}}}\right)_{\varepsilon}
&=&{\frac{\mathrm{d}\Gamma}{\mathrm{d}T_{\mathrm{E}}}}\,
\cdot\,{\cal E}(T_{\mathrm{E}}) = {\frac{32\, G^{2} \,
\mathrm{M}_{\sun}\, {\cal E}(T_{\mathrm{E}})}{c^{4}\,
 T_{\mathrm{E}}^{4}\, {\cal N}}}\,
\,\cdot \nonumber\\
&&
\int_{d_{\mathrm{min}}}^{d_{\mathrm{max}}}\mathrm{d}D_{\mathrm{os}}
\,D_{\mathrm{os}}^{3}\, \cdot \nonumber\\
&&
\int_{\frac{d_{\mathrm{min}}}{D_{\mathrm{os}}}}^{1}\mathrm{d}
x\, x^{2}\, (1-x)^{2} \int_{0.08}^{1.5} \mathrm{d} \mu\, \mu^{2}\,
{\frac{\mathrm{d}
n_{0}}{\mathrm{d} \mu}} \cdot \nonumber\\
&&
\left[{\frac{\rho_{\mathrm{b}}(x)\rho_{\mathrm{b}}(D_{\mathrm{os}})}
{(1+x)^{2}\sigma_{\mathrm{b}}^{2}}}
\,e^{\left(-{\frac{2\, G\, \mathrm{M}_{\sun}\, D_{\mathrm{os}}\,
x\, (1-x)\, \mu}{ c^{2}\, (1+x^{2})\, \sigma_{\mathrm{b}}^{2}\,
T_{\mathrm{E}}^{2}}}\right)}\right. + \nonumber\\
&& {\frac{\rho_{\mathrm{d}}(x)\rho_{\mathrm{b}}(D_{\mathrm{os}})}
{\sigma_{\mathrm{d}}^{2}+x^{2}\sigma_{\mathrm{b}}^{2}}}
\,e^{\left(-{\frac{2\, G\, \mathrm{M}_{\sun}\, D_{\mathrm{os}}\,
x\, (1-x)\, \mu}{ c^{2}\,
(\sigma_{\mathrm{d}}^{2}+x^{2}\sigma_{\mathrm{b}}^{2})\,
T_{\mathrm{E}}^{2}}}\right)}+ \nonumber\\
&& {\frac{\rho_{\mathrm{b}}(x)\rho_{\mathrm{d}}(D_{\mathrm{os}})}
{\sigma_{\mathrm{b}}^{2}+x^{2}\sigma_{\mathrm{d}}^{2}}}
\,e^{\left(-{\frac{2\, G\, \mathrm{M}_{\sun}\, D_{\mathrm{os}}\,
x\, (1-x)\, \mu}{ c^{2}\,
(\sigma_{\mathrm{b}}^{2}+x^{2}\sigma_{\mathrm{d}}^{2})\,
T_{\mathrm{E}}^{2}}}\right)}+ \nonumber\\
&&
\left.{\frac{\rho_{\mathrm{d}}(x)\rho_{\mathrm{d}}(D_{\mathrm{os}})}{(1+x)^{2}\sigma_{\mathrm{d}}^{2}}}
\,e^{\left(-{\frac{2\, G\, \mathrm{M}_{\sun}\, D_{\mathrm{os}}\,
x\, (1-x)\, \mu}{ c^{2}\, (1+x^{2})\, \sigma_{\mathrm{d}}^{2}\,
T_{\mathrm{E}}^{2}}}\right)}\right] \label{deGammaSuDeT}
\end{eqnarray}

In the following section we need also the two distributions
\begin{equation}
\left({\frac{\mathrm{d}^{2}\Gamma}{\mathrm{d}T_{\mathrm{E}}\,
\mathrm{d}x}}\right)_{\varepsilon} =\left. {\cal
E}(T_{\mathrm{E}})\,{\frac{\mathrm{d}^{2}\Gamma}
{\mathrm{d}T_{\mathrm{E}}\,\mathrm{d}x}}\right|_{T_{\mathrm{E,event}}}
\end{equation}
\begin{equation}
\left({\frac{\mathrm{d}^{2}\Gamma}{\mathrm{d}T_{\mathrm{E}}\,
\mathrm{d}\mu}}\right)_{\varepsilon} =\left. {\cal
E}(T_{\mathrm{E}})\,{\frac{\mathrm{d}^{2}\,\Gamma}{\mathrm{d}
\,T_{\mathrm{E}}\,\mathrm{d}\mu}}\right|_{T_{\mathrm{E,event}}}
\end{equation}
calculated assigning to $T_{\mathrm{E}}$ the corresponding
effective measured value.

%
\subsection{A statistical analysis for self--lensing events
di\-scri\-mi\-na\-tion} \label{sec:gtestat}
In this section we will show that, in the framework of the LMC
geometrical structure and dynamics outlined in Sect. \ref{sec:modelli}, a
suitable statistical analysis allows us to exclude from the
self--lensing population a large subset of the detected events. To
this purpose, assuming  all the 15 events as self lensing, we
study the  scatter plots  correlating the self lensing expected
values of some meaningful microlensing variables with the measured
Einstein time or with the self--lensing optical depth. The
idea underlying this analysis is based on the search of self--lensing 
average collective properties with different behaviour in
the two different regions of the LMC: the bar with its nearby
neighbourhood and the disk region external to it. In this way we
eventually show that a large subset of events is
incompatible with the self--lensing hypothesis.

\begin{figure}
\resizebox{\hsize}{!}{\includegraphics{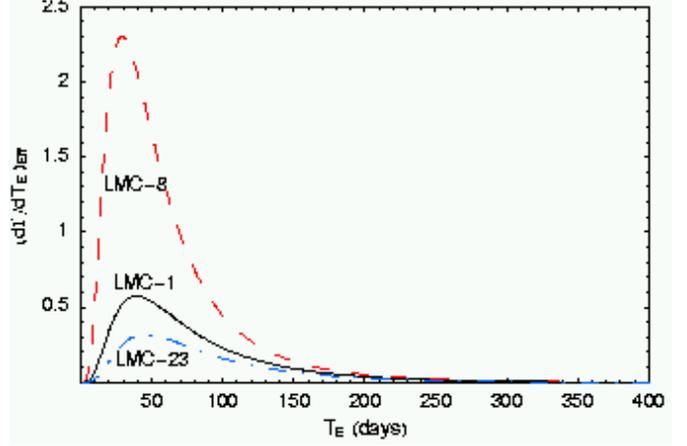}}
\caption{Differential rate of the microlensing events with respect to the
Einstein time $T_{\mathrm{E}}$, along the lines of sight pointing
towards the events LMC--8 (dashed red line), LMC--1 (solid black
line) and LMC--23 (dot dashed blue line). The $y$--axis values are
in $10^{-12}$ units.} \label{fig:eventi1e8e23}
\end{figure}

We have calculated the self--lensing distributions
$\left({\frac{\mathrm{d}\Gamma}{\mathrm{d}T_{\mathrm{E}}}}\right)_{\varepsilon}$
of the rate of microlensing events with respect to the Einstein
time $T_{\mathrm{E}}$, along the lines of sight towards the 15
events of the MACHO collaboration, in the case of a Chabrier
exponential type IMF. As an example we show in Fig.
\ref{fig:eventi1e8e23} the distributions
$\left({\frac{\mathrm{d}\Gamma}{\mathrm{d}T_{\mathrm{E}}}}\right)_{\varepsilon}$
calculated along the lines of sight pointing towards the events
LMC--1 (solid line), LMC--8 (dashed line) and LMC--23 (dot dashed
line)\footnote{We have also used the Chabrier power law IMF,
obtaining slightly higher values, of the order of 10\%, with
respect to the exponential one. In the following we have used
constantly the Chabrier exponential IMF.}.

With these distributions we have calculated the median\footnote{As
the distribution is strongly asymmetric, to describe the expected
value of $T_\mathrm{E}$ we use the median value, a more meaningful
parameter than the average value.} $T_{\mathrm{E},\,50\,\%}$ and
the values $T_{\mathrm{E},\,16\,\%}$ and $T_{\mathrm{E},\,84\,\%}$
that single out the extremes of the 68\% probability range around
the median (not to be confused with a 1$\sigma$ error). In
Table \ref{tab:gtecomp} we report these values for each observed
MACHO event.
\begin{table*}
\centering
\begin{tabular}[h]{ccccccc}
\hline
\hline
event & x (kpc) & y (kpc) & $T_{\mathrm{E,obs}}$ & $T_{\mathrm{E,50\%}}$
& $T_{\mathrm{E,16\%}}$ & $T_{\mathrm{E,84\%}}$ \\
\hline
 1 & 1.017 &  0.909 & 22.3  & 64 & 33 & 126\\ \hline
 4 & 0.746 & -0.814 & 29.5  & 66 & 35 & 128\\ \hline
 5 & 0.797 & -0.559 & 49.1  & 65 & 34 & 127\\ \hline
 6 & 0.102 & -0.423 & 59.5  & 55 & 28 & 111\\ \hline
 7 & 1.796 & 0.189 & 66.8  & 73 & 38 & 140\\ \hline
 8 & 0.185 &  0.062 & 43.1  & 48 & 25 & 95\\ \hline
13 & 0.280 &  0.916 & 65.0  & 66 & 33 & 128\\ \hline
14 & -0.523 & -0.487 & 65.0  & 51 & 26 & 103\\ \hline
15 & 1.652 &  0.048 & 23.9  & 72 & 38 & 137\\ \hline
18 & -1.253 & -1.168 & 48.2  & 72 & 38 & 138\\ \hline
20 & 2.478 & -0.316 & 47.2  & 77 & 40 & 146\\ \hline
21 & 2.324 & 0.211 & 60.5  & 77 & 40 & 146\\ \hline
23 & 1.517 & -1.037 & 55.4  & 72 & 38 & 138\\ \hline
25 & 2.072 & 1.517 & 55.4  & 79 & 41 & 149\\ \hline
27 & 1.619 & 0.388 & 32.8  & 67 & 34 & 130\\ \hline
\end{tabular}
\caption{Microlensing rate: the results for the predicted Einstein time.
Case with disk and bar coplanar and bar velocity
dispersion $\sigma_{\mathrm{b}} = 20.2$ km/s. For each observed
MACHO event we report for the Einstein time (days): the observed
value and the predicted median value with the two extreme of a 68\%
probability range around it. In the second and third column we
report the position in the reference frame centered in the LMC
as defined in Sect. \ref{sec:modelli}.} \label{tab:gtecomp}
\end{table*}

Trying to see if the geometry can help, we have analyzed how the
self--lensing expected values  of $T_\mathrm{E}$ depend on the
position, or better still on the optical depth, taking into
account that the LMC disk symmetry is elliptical and not circular.

\begin{figure}
\resizebox{\hsize}{!} {\includegraphics{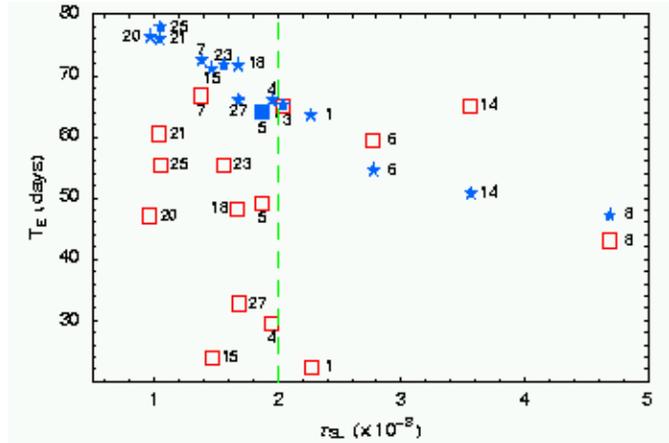}}
{\caption{Scatter plot of the observed (empty boxes) values of the
Einstein time and of the expected values of the median
$T_{\mathrm{E},50\, \%}$ (filled stars), with respect to the
self--lensing optical depth evaluated along the directions of the
events. The $T_{\mathrm{E},50\, \%}$ for the galactic disk 
event LMC--5 is represented by a
filled box.} \label{fig:tevstau}}
\end{figure}

In Fig. \ref{fig:tevstau} we report on the $y$--axis the observed
values of $T_\mathrm{E}$ (empty boxes) as well as the expected
values for self lensing of the \emph{median}
$T_{\mathrm{E}\,,50\,\%}$ (filled circle) evaluated \emph{along
the directions of the events}. On the $x$--axis we report the
value of the self--lensing optical depth calculated towards the
event position. The optical depth increases as one moves from the outer
regions towards the center of the LMC according to the contour lines
shown in Fig. \ref{fig:SL}. An interesting feature emerging clearly is
the \emph{decreasing} trend of the  expected values of the median
$T_{\mathrm{E}\,,50\,\%}$, going from the outside fields with low
values of $\tau_{\mathrm{SL}}$ towards the central fields with
higher values of $\tau_{\mathrm{SL}}$. The variation of the
stellar number density and the flaring of the LMC disk certainly
contributes to explain this behaviour.

We now tentatively identify two subsets of events: the ten falling
outside the contour line $\tau_{\mathrm{SL}} = 2 \times 10^{-8}$
of Fig. \ref{fig:SL} and the five falling inside. In the framework of
van der Marel et al. LMC geometry, this  contour line includes
almost fully the LMC bar and two ear shaped inner regions of the
disk, where we expect self--lensing events to be located with
higher probability.

At first glance, we note that the two clusters have a clear-cut
different collective behaviour: the measured Einstein times of the
first 10 points fluctuate around a median value of 47 days, very
far from the expected values of the median $T_\mathrm{E}$, ranging
from 65 days to 79 days, with an average value of 72 days. On the
contrary, for the last 5 points, the measured Einstein times
fluctuate around a median value of 51 days, near to the average
value 57 days of the expected medians, ranging from 48 days to 66
days. Let us note, also, the somewhat peculiar position of the
event LMC--1, with a very low value of the observed
$T_{\mathrm{E}}$. In the following analysis it will be shown that
most probably this event is homogeneous to the set at left of the
vertical line in Fig. \ref{fig:tevstau} and it has to be included in
that cluster.

This plot gives a first clear evidence that, in the framework of
van der Marel et al. LMC geometry, the self--lensing events have
to be searched among the cluster of events with
$\tau_{\mathrm{SL}}\,>\, 2\times 10^{-8}$, and at the same time
that the cluster of the $10$ events plus LMC--1  belongs, very
probably, to a different population.

\begin{figure}
\resizebox{\hsize}{!}{\includegraphics{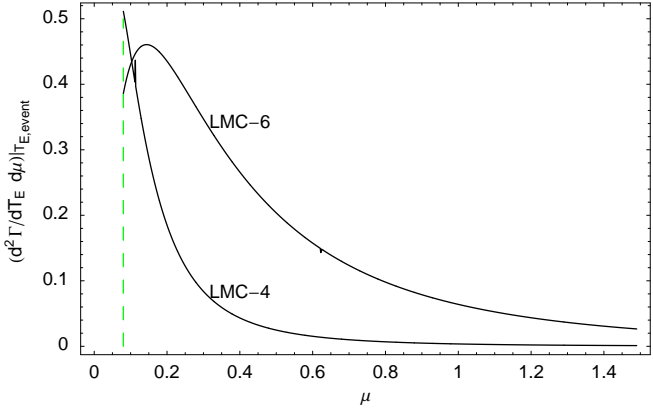}}
\caption{Differential rate of the microlensing events with respect
to the mass of the lens in direction of events LMC--4 and LMC--6.
The dashed vertical line is at $\mu=0.08$. The y axis values are in
$10^{-3}$ units.} \label{fig:deGammaVersusDeMu}
\end{figure}
\begin{figure}
\resizebox{\hsize}{!}{\includegraphics{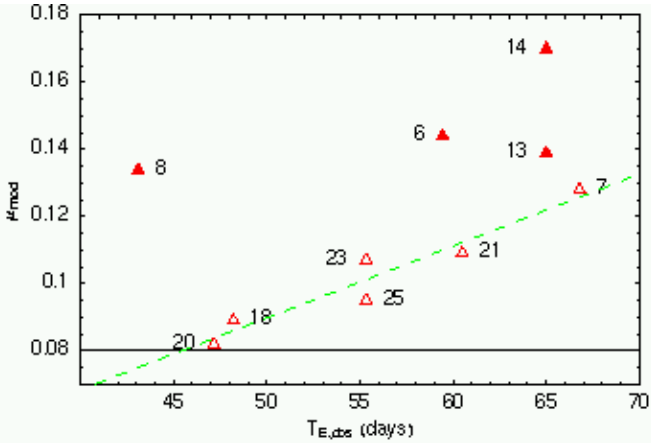}}
\caption{Scatter plot of the modal value of $\mu$ ($y$--axis) with
respect to the measured Einstein time ($x$--axis) of each event.
The label identifies the event. Filled (empty) triangles represent
points with $\tau_{\mathrm{SL}}\, >\, 2\times 10^{-8}\,(<\,2\times
10^{-8})$. The horizontal line indicates the lower limit of the
lens mass for self lensing, $\mu=0.08$. The four events LMC--1,
LMC--4, LMC--15 and LMC---27 have a modal value of the lens mass
smaller than the lower limit and therefore are not represented.
The dashed line represents the correlation line for the $6$ events
with $\tau_{\mathrm{SL}}\, <\, 2\times 10^{-8}$.}
\label{fig:tEVersusMu}
\end{figure}

In order to further improve our statistical analysis, we have
calculated the distributions
$\left({\frac{\mathrm{d}^{2}\Gamma}{\mathrm{d}T_{\mathrm{E}}\,
\mathrm{d}\mu}}\right)_{\varepsilon}$ along the lines of sight
pointing towards the 14 LMC microlensing events\footnote{The
previous 15 events minus the event LMC--5, formerly
recognized as a Galactic disk event.}, taking, for each line of
sight, the observed Einstein time value. As an example we show in
Fig. \ref{fig:deGammaVersusDeMu} this distribution calculated for the
two events LMC--6 and LMC--4. LMC--6 is representative of the
events for which the modal value $\mu_{\mathrm{mod}}$ falls inside
the self--lensing mass interval 0.08--1.5. The second one has been
chosen to put in evidence that there are also events for which
$\mu_{\mathrm{mod}}$ falls outside, in the range 0--0.08.

Fig. \ref{fig:tEVersusMu} is the scatter plot between the measured
$T_{\mathrm{E}}$ ($x$--axis) and the modal value 
of the lens mass $\mu={\frac{M_{\mathrm{l}}}{\mathrm{M}_{\sun}}}$
($y$--axis), calculated by the distribution
$\left({\frac{\mathrm{d}^{2}\Gamma}{\mathrm{d}T_{\mathrm{E}}
\,\mathrm{d}\mu}}\right)_{\varepsilon}$. In this case we prefer to
use the modal value rather than the median one, because it is
independent of the amplitude of the interval of the allowed
values of the lenses mass, whereas the median value varies in
accordance with this choice.

We find that the events LMC--1, LMC--4, LMC--15 and LMC--27 have a
modal value of the lens mass smaller than the lower limit. We
consider this result as a strong indication to exclude these four
events from the self--lensing class. We have then
calculated the linear correlation between $\mu_{\mathrm{mod}}$ and
$T_{\mathrm{E}}$ for the 6 remaining points out of the cluster of
ten. We find a high linear correlation, as shown by the dashed
green straight line in Fig. \ref{fig:tEVersusMu} and by the calculated
linear correlation coefficient equal to $0.963$. We observe that
the values of $\mu_{\mathrm{mod}}$ of the six events range between
0.08 and 0.13, an interval narrow enough to justify the linear
approximation to represent a small portion of a parabolic curve,
around which we expect the correlated points to disperse, bearing
in mind that $T_{\mathrm{E}}$ is proportional to $\sqrt{\mu}$. The
six linearly correlated events, therefore, constitute a
homogeneous population, clearly distinct, also in this parameter
space, from that formed by events with $\tau_{\mathrm{SL}}\,>\,
2\times 10^{-8}$ for which, for a given value of the observed
Einstein time, we get significantly higher values for the mass.

\begin{figure}
\resizebox{\hsize}{!}{\includegraphics{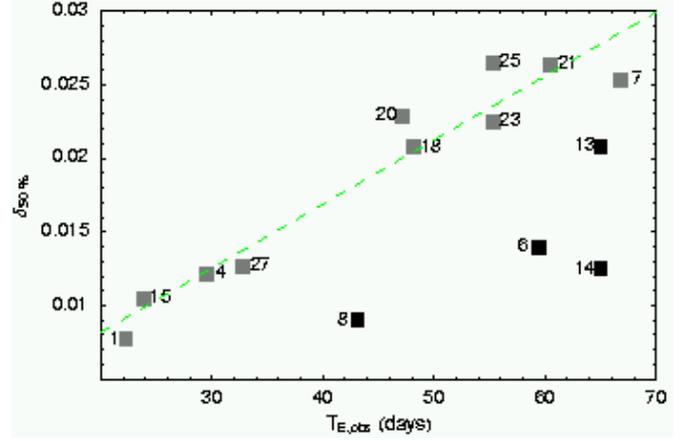}}
\caption{Correlation plot of the median value (square boxes) of
$\delta={\frac{D_{\mathrm{os}}-D_{\mathrm{ol}}}{D_{\mathrm{os}}}}$
($y$--axis) with respect to the measured Einstein time ($x$--axis)
of each event. The label identifies the event. The dashed line is
the correlation straight line of the cluster of 10 events.}
\label{fig:tEVersusX}
\end{figure}

We have calculated also the distributions
$\left({\frac{\mathrm{d}^{2}\Gamma}{\mathrm{d}T_{\mathrm{E}}\,\mathrm{d}
x}}\right)_{\varepsilon}$ along the lines of sight pointing towards
the 14 LMC microlensing events, in order to obtain a second
independent check of the homogeneity of the cluster of 
events with $\tau_{SL}\le 2\times 10^{-8}$. Fig. \ref{fig:tEVersusX} is the scatter plot between the
measured $T_{E}$ (x-axis) and the median (square boxes) of the
parameter
$\delta={\frac{D_{\mathrm{os}}-D_{\mathrm{ol}}}{D_{\mathrm{os}}}}=1-x$
($y$--axis), proportional to the distance between the lens and the
source.

We find that the  points of the first cluster  have a high
linear correlation, as shown by the calculated linear correlation
coefficient equal to 0.955. We observe also that adding the event
LMC--1 to the set of 9 and recalculating the linear correlation
coefficient between $\delta$ and $T_{\mathrm{E}}$, we find a small
increase, from 0.955 to 0.965, suggesting that this event, lying
inside the contour line at optical depth
$\tau_{\mathrm{SL}}=2\times 10^{-8}$, is very probably homogeneous
to the population of the 9 falling in the outer part. The ten
events, represented by gray square boxes, are highly correlated as
shown by the dashed green straight line in Fig. \ref{fig:tEVersusX}.
This is, again, a strong indication that they constitute an
homogeneous population of events. Together with the fact that, as
shown in Fig. \ref{fig:tevstau}, the measured Einstein time fluctuates
around a median value very far from the median Einstein times,
calculated with the self--lensing formulae, this allows us to
exclude that this class of events belong to self lensing. But,
before any definitive assessment on the nature of these events can
be made, we have to wait for an analogous statistical analysis for
the case of microlensing events due to lenses in the halos of the LMC
or of our Galaxy: such an analysis is now in preparation for a
forthcoming paper.

We come now to the discussion of the changes in the
microlensing rate induced by the different bar geometry
configurations and velocity dispersions, as introduced in Sect.
\ref{sec:bar}. The main point to be stressed is that the separation in the
two populations for the events already identified is in these
cases enhanced. Indeed, the expected characteristics change
significantly only for those events along the lines of sight
pointing towards the central region: LMC--6, LMC--8 and LMC--14, and,
marginally, LMC--1 and LMC--27, as it is the case for the self--lensing
optical depth. In Table \ref{tab:gtenew} we report the predicted
values for the Einstein time for these 5 events, the changes for
the others being at most of $\sim$ 5\% (in particular this is the
case for the event LMC--13). 
\begin{table*}
\centering
\begin{tabular}[h]{cccccccc}
\hline
\hline
event & $T_{\mathrm{E,obs}}$ & $T_{\mathrm{E,50\%}}$  &
$T_{\mathrm{E,50\%}}$  & $T_{\mathrm{E,50\%}}$   &
$\tau\,(10^{-8})$ & $\tau\,(10^{-8})$ & $\tau\,(10^{-8})$ \\
& & coplanar & rotated & shifted &coplanar & rotated & shifted \\
 \hline
  1 & 22.3  & $64\quad 56$ & $76\quad 67$ & $64\quad 56$ & 2.24 & 3.33 & 2.23\\ \hline
  6 & 59.5  & $55\quad 47$ & $57\quad 48$ & $62\quad 53$ & 2.84 & 3.16 & 3.90\\ \hline
  8 & 43.1  & $48\quad 37$ & $53\quad 41$ & $51\quad 40$ & 4.72 & 5.99 & 5.58\\ \hline
 14 & 65.0  & $51\quad 41$ & $60\quad 47$ & $56\quad 44$ & 3.72 & 5.33 & 4.53\\ \hline
 27 & 32.8  & $67\quad 59$ & $75\quad 67$ & $69\quad 61$ & 1.75 & 2.29 & 1.94\\ \hline
\hline
\end{tabular}
\caption{For each MACHO event located within or nearby the bar
region, where the variation due to the change in the geometry turn
out to be significant, we report for the Einstein time (days): the
observed value and the predicted values for the three geometry
configurations, bar coplanar with the disk, bar rotated by
$40^{\circ}$ with respect to the disk plane and bar shifted by 0.5
kpc along the line of sight towards the observer. The left values
are for a velocity dispersion of the bar component of
$\sigma_{\mathrm{b}}$=20.2 km/s (as for the disk), whereas the
right values are for $\sigma_{\mathrm{b}}$ = 30.0 km/s. We report
also the values for the self--lensing optical depth in the
direction of the events.} \label{tab:gtenew}
\end{table*}
As expected, an increase in the
velocity dispersion for the bar component leads
to a decrease ($\sim$ 20\%) for the predicted values of the
Einstein time. For the events LMC--6, LMC--8 and LMC--14 the predicted
values just fluctuate around the observed ones, while they remain
substantially different for the events LMC--1 and LMC--27. The change
in geometry leads, on the contrary, to a rise for the expected
values of the Einstein time, at most $\sim$ 10\% (the effect being
stronger for the rotated bar), for the events LMC--6, LMC--8 and LMC--14. For
the event LMC--1 the difference between the observed and the
predicted value is further enhanced (up to $\sim$ 16\%). This
gives further support to the guess that this short duration event
is likely to belong to a different population. We note that the
two opposite effects on the predicted Einstein time, the decrease
linked to the rise of the velocity dispersion opposed to the
increase linked to the non coplanar geometry, are actually to be
expected as $\Gamma \propto \tau/T$. Analogous considerations
emerge from the analysis of the differential rate with respect to
$x$ and the mass. In particular, for the mass, we note again
two opposite effects. A rise induced by the increase in the
velocity dispersion, and a decrease linked to the non coplanar
geometry. We observe  a maximum increase of $\sim 15$\% for 
the event LMC--14.

To complete this analysis let us make a further point. 
In the differential rate, Eq. \ref{deGammaSuDeT},
we get an average result for all the possible configurations
with lens and source either in the bar or in the disk of the LMC.
This is of course coherent with our approach aimed at the study
of the statistical properties of the self lensing
population as a all. However, it is in principle interesting
to notice that the expected characteristics can be 
actually rather different depending on the configuration.
As an example, we consider the event LMC--14, coplanar
bar geometry, $\sigma_\mathrm{b}=20.2$ km/s. 
With respect to the previously obtained result, expected median value
for the  Einstein time $T_{\mathrm{E},50\%}$=51 days,
we get a shorter value in the lens-source bar-bar configuration,
$\sim 43$ days, a significantly longer value for the disk-disk configuration,
$\sim 67$ days, and an intermediate one, $\sim 55$ days, for both
mixed configurations, disk-bar and bar-disk. We recall that the observed
value for LMC--14 is $T_{\mathrm{E}}$=65 days.

\subsection{Expected number of self--lensing events} \label{sec:nevsl}
We compute the ``field exposure'', $E_{\mathrm{field}}$, defined,
as in \citet{alcock00a}, as the product of the number of distinct
light curves per field by the relevant time span, paying attention
to eliminate the field overlaps. Furthermore we calculate the
distribution ${\frac{\mathrm{d}\Gamma}{\mathrm{d}T_{\mathrm{E}}}}$
along the line of sight pointing towards the center of each field.
In this way we obtain the number of  expected events for
self lensing, field by field, given by
\begin{equation}
N_{\mathrm{SL,field}}=E_{\mathrm{field}}\int_{0}^{\infty}\,
{\frac{\mathrm{d}\Gamma}{\mathrm{d}T_{\mathrm{E}}}}\, {\cal
E}(T_{\mathrm{E}})\, \mathrm{d}\,T_{\mathrm{E}} \; ,
\end{equation}
where ${\cal E}(T_{\mathrm{E}})$ is the detection efficiency.

In Fig. \ref{fig:asymmetry} we report in parenthesis the expected
number of events in each field we obtain by using the Chabrier
exponential IMF. Summing over all fields we find that the expected
total number of self--lensing events is $\sim 1.2$, while we would
get $\sim 1.3$ with the the double power law IMF, in both cases
altogether 1-2 events. Clearly, taking also into account the
uncertainties in the parameter used following the van der Marel
model for the LMC the actual number could also be somewhat higher
but hardly more than twice our estimate. A conclusion we had
already reached in Paper I.

We have also computed the influence of the bar geometry and
dispersion velocity on the number of expected events. We find at
most an increase of $\sim$ 50\%, bar rotated and
$\sigma_{\mathrm{b}}$= 30 km/s, that gives altogether about 2-3
events.

In conclusion, we have identified, according to their
expected characteristics as compared to the observed one, a set of
4 possible events belonging to the self--lensing population
(LMC--6, LMC--8, LMC--13 and LMC--14). As already observed, there are some
experimental evidences that the LMC--14 is a self--lensing event 
\citep{alcock01b}.
If this is true, our prediction of at most 2-3 such events seems
thus to indicate that some of the others are more likely not to be
due to self lensing.

\section{Asymmetry} \label{sec:asimmetria}
In this section we consider from a different perspective the issue
of the expected near/far asymmetry for events due to lenses
located in the LMC or the MW halo. In particular we study, by
means of a statistical analysis, the spatial distribution of the
events, both in the framework of the van der Marel picture and,
for comparison, of the older one. The aim here is to give further
support to our main conclusion, namely that self lensing, for
which we do not expect any asymmetry in the spatial distribution,
can not explain all the detected events.
\begin{figure*}
 \resizebox{\hsize}{!}{\includegraphics{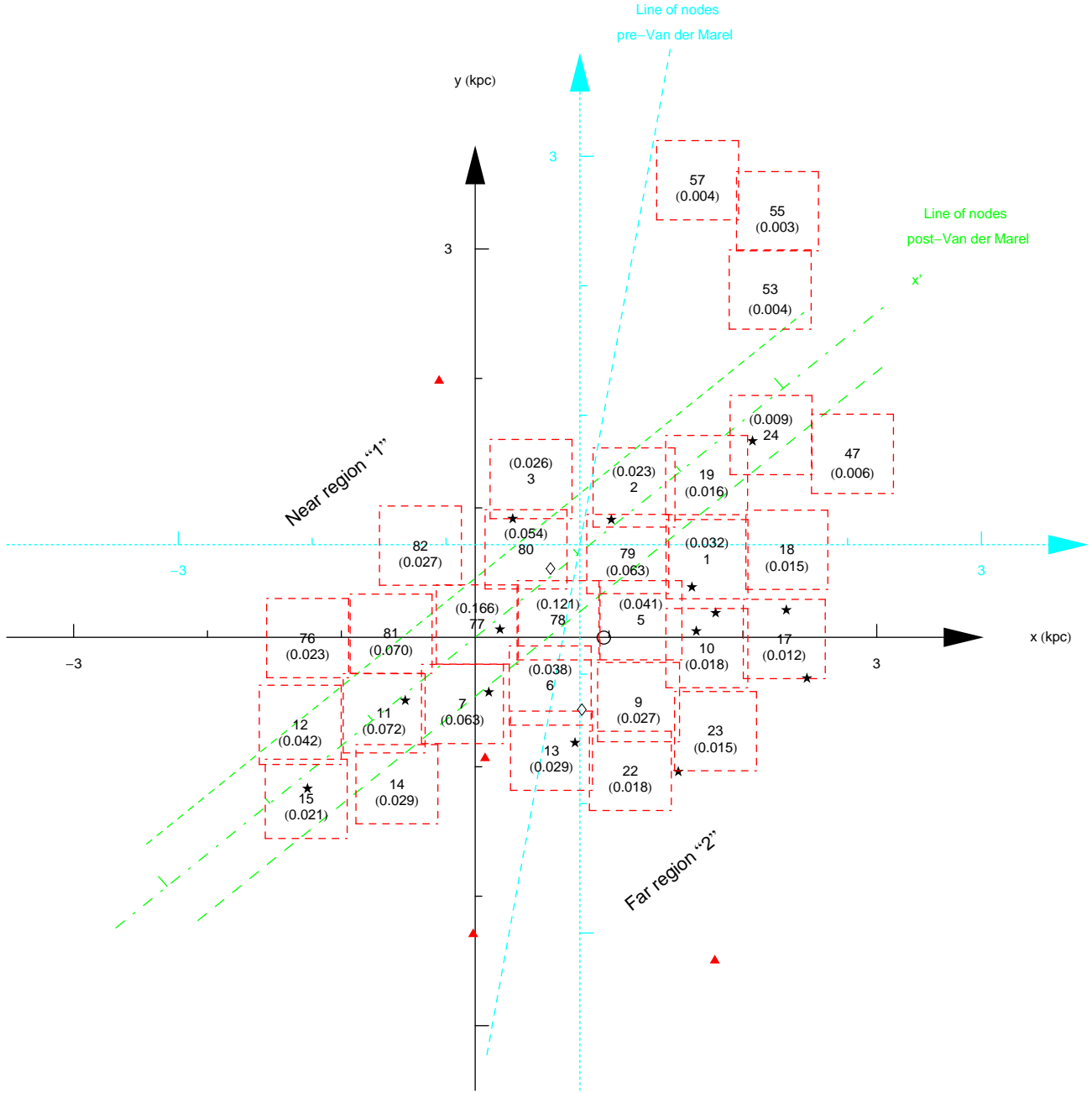}}
 \caption{Locations of the 30 MACHO fields with respect to post-Van
 der Marel LMC geometry (black reference axes) and with respect to
 pre-Van der Marel LMC geometry (light blue reference axes).
 Also shown are the locations of the MACHO and EROS microlensing candidates.
 The distances on the axes are in kpc. For each field, inside brackets,
 we report the expected number of self--lensing event as discussed in Sect. \ref{sec:nevsl}.}
 \label{fig:asymmetry}
\end{figure*}

In Fig.  \ref{fig:asymmetry}, the 30 well-sampled fields analyzed by
the MACHO collaboration (red squares), together with the 16
events\footnote{The ensemble of 17 events discussed in Sect.
\ref{sec:mlpar} with the exclusion of LMC-22.}, are located in a
reference frame (black axes) centered in $\alpha_{\mathrm{CM}} =
5^{\mathrm{h}} 27.6^{\mathrm{m}}$, $\delta_{\mathrm{CM}} = - \,
69.87^{\circ}$, J$2000$. We also report the position (triangles)
of the microlensing events found by the EROS collaboration
\citep{lasserre00} even if we do not consider them in the present
analysis.

We divide the LMC field in three regions: a strip centered on the
line of nodes and delimited by two parallel straight lines  at a
distance of $\approx 0.35\, \mathrm{kpc}$ from the line of nodes,
and two outer regions, belonging to the closer north-east side and
to the farther south-west one.  The amplitude of the central strip
reflects the uncertainties in the position of the center of mass
of carbon stars, as calculated by \citet{marel02}, at one $\sigma$ level.
The exclusion of a substantial
part of the bar region implies that the following discussion is
not affected by the different choices for the bar geometry.

The green dashed--dot line represents the line of nodes with a
position angle $\Theta = 129.9^{\circ}$ and the two green dashed
lines delimit the two outer regions  belonging respectively to the
near and far part of the LMC.

On the same figure a second reference frame is also reported,
according to  pre-van der Marel LMC models. The frame (light blue
axes) is centered in $\alpha_{\mathrm{gas}} = 5^{\mathrm{h}}
17.6^{\mathrm{m}}$, $\delta_{\mathrm{gas}} = - \, 69^{\circ} 02'$,
J2000 \citep{kim98}, and the line of nodes has a position angle
$\Theta_{\mathrm{Gyuk}} = 170^{\circ}$ \citep{gyuk00}. A region of
exclusion, similar to the previous one will be considered also in
this case, even if not drawn on the figure for clarity sake.

At first glance one observes that the distribution of the events
shows a clear near-far asymmetry in the post-van der Marel
geometry, namely, they are concentrated along the extension of the bar and
in the south-west side of the LMC. On the contrary, the asymmetry is
almost completely lost in the pre-van der Marel geometry, where
the distribution reflects almost exactly the different weights of
the observed fields in the two half planes, as we will show in the
following.

The little empty circle in the field number 5 locates the position
of the baricenter of the 16 MACHO events, whose coordinates in the
post-van der Marel reference frame are (0.96 kpc, -- 0.02 kpc).

\subsection{Near-far asymmetry of the observed microlensing events in the LMC}
In order to give a quantitative analysis of the near--far
asymmetry, we will concentrate on the MACHO observed fields (and
the corresponding observed events) in the outer regions located on
the opposite sides of the line of the nodes and external to  the
dashed lines, which, at a good confidence level, belong
respectively to the near and far part of the galaxy.

As a first point we determine the fraction of the MACHO fields
included respectively in the north-east closer region ``1'' and
in the south-west farther  region ``2'', and we calculate in each
the quantities $\Sigma_{1}$ and $\Sigma_{2}$, defined as the sum
of the product of star number pro field, times the corresponding
observation time, where we count only once the stars in the intersecting
part of the fields. The ratio $\Sigma_{i}/(\Sigma_{1}+\Sigma_{2})$
gives us the probability $p_{i}$ that a microlensing event would
fall in the first or second region, respectively. Let us note that
the probability scheme so defined depends only on the global
observation strategy in the near and far region (fields
distribution and  observation time). We find $p_{1} \approx 0.31$
in the case of the post-van der Marel LMC geometry, and
$\tilde{p}_{1} \approx 0.48$ for the pre-van der Marel LMC
geometry.

We are interested in testing whether the observed events support
the modelled probability schemes. The Pearson chi-square
statistic ${\cal  A}_{N}$ provides a non parametric test for the
comparison:
\begin{equation}
{\cal  A}_{N}= \sum_{i=1}^{2} {{(n_{i}-N p_{i})^{2}}\over N p_{i}}~,
\end{equation}
where $n_{i}$ are the events falling in the $i$-region, and $N=
n_{1}+n_{2}$. ${\cal  A}_{N}$ is approximately distributed as
chi-square with one degree of freedom. Small values of ${\cal
A}_{N}$ tend to support the null hypothesis that the $p_{i}$
match the measured distribution.

In the post-van der Marel geometry we get, respectively, 1 and
9 observed events in regions ``1'' and ``2'', and then
$${{\cal  A}_{10}} = 2.11.$$
On the contrary, in the case of pre-van der Marel geometry, we obtain:
$${\tilde{\cal A}_{12}} = 0.22,$$
starting from a set of 5 and 7 observed events.

In the pre-van der Marel geometry ${{\cal A}_{N}}$ gets a value
near to zero. At the confidence level of $\simeq 64 \%$ the null
hypothesis  that the events distribution reflects almost exactly
the weights of the two regions has to be accepted. This implies
that the distribution of the lenses should be almost homogeneous
along the lines of sight through the different regions of the LMC.

On the contrary, in the post-van der Marel geometry ${{\cal
A}_{N}}$ assumes a value far enough from zero. At the confidence
level of $\simeq 85\%$ the null hypothesis must be rejected. This
means that the observed asymmetry is greater than the one expected
simply on the basis of the observational strategy.

We are aware that these results have to be treated with caution
inasmuch as the number of events is small. However, we note that
the observed near--far asymmetry is coherent with that induced by
the inclination of the LMC disk already discussed, looking at the
optical depth, in Sect. \ref{sec:tau}.

\section{An improved Gould inequality} \label{sec:gould}
\citet{gould95} ingenious calculation provides two inequalities
(Eq. (2.8) and Eq. (3.3) of his paper), the second one involving
the Jeans equation and the virial theorem, which allow, in some
cases, a quick evaluation of an upper limit to the self--lensing
optical depth along a line of view through the center of a galaxy.
In particular this was applied to the LMC disk, in order to exclude
the \citet{sahu94} hypothesis that the observed optical depth
could be fully explained by self lensing.

Our aim here is to obtain an improved version of Gould
inequality, and, at the same time, clarify its limits of
applicability. We start from Eq. (\ref{weightedOD}) and consider
the case of a line of view passing through the center of the LMC.

Let us assume that the lens mass density $\rho_{\mathrm{l}}$ be a
function of an {\it homogeneous polynomial of degree}
$n_{\mathrm{l}}$ in the variables $(\xi,\,\eta,\,\zeta)$ of the
reference frame defined in Sect. \ref{sec:disk}, and analogously that the
star mass density $\rho_{\mathrm{s}}$ be a function of an {\it
homogeneous polynomial of degree} $n_{\mathrm{s}}$\footnote{This
is implicit in the \citet{gould95} derivation.}. Let us also
denote with $z$ the current coordinate along the line of view
through the center of the LMC disk, which we assume as the origin of
the $z$--coordinate, and with $i$ the disk inclination angle.
Keeping in mind that for points belonging to a line of view
passing through the center it results:
\begin{eqnarray}
\xi=& 0\nonumber\\
\eta=& z \sin i  \nonumber\\
\zeta=& z \cos i \nonumber
\end{eqnarray}
we can write for the lens density distribution
\begin{eqnarray}
\rho_{l}[0,\,\eta,\,\zeta]& =
\rho_{l}\left[\sum_{k=0}^{n}\,c_{k,n-k}^{(\mathrm{l})}
\,\left({\frac{\eta}{\eta_{\mathrm{l},0}}}\right)^{k}
\,\left({\frac{\zeta}{\zeta_{\mathrm{l},0}}}\right)^{n-k}\right] \nonumber\\
&= \rho_{\mathrm{l}}\left[\gamma_{\mathrm{l}}(i)\,
\left({\frac{z}{\zeta_{\mathrm{l},0}}}\right)^{n}\right]
\end{eqnarray}
with
\begin{equation}
\gamma_{\mathrm{l}}(i) =
\sum_{k=0}^{n}\,c_{k,n-k}^{(\mathrm{l})}\,
\left({\frac{\eta_{\mathrm{l,}0}}{\zeta_{\mathrm{l},0}}}
\right)^{k}\sin^{k}(i)\,\cos^{n-k}(i)
\end{equation}
and $\eta_{\mathrm{l},0}$, $\zeta_{\mathrm{l},0}$ the scale
lengths of the lenses density distribution. Analogous expressions
can be written for the star density distribution, changing
properly the suffix ``l'' with ``s''.

From now on we follow \citet{gould95}, and put $D_{\mathrm{ol}}=
d+z$ and $D_{\mathrm{os}}= d+w$, where $d$ is the distance, along
the line of view, from the observer to the center of the disk.
Taking into account that
$${\frac{d+z}{d+w}}<1,$$
we obtain the following inequality for the optical depth:
\begin{eqnarray}\label{inequality}
\tau \le &{\frac{4\pi G}{c^{2}}}\int_{-t}^{t} \mathrm{d}w
\,\rho_{\mathrm{s}}\left[\gamma_{\mathrm{s}}(i)\,\left({\frac{w}{\zeta_{\mathrm{s},0}}}
\right)^{n_\mathrm{{\mathrm{s}}}}\right]
\cdot \nonumber\\
&\cdot{\frac{\int_{-d}^{w} \mathrm{d}z\,(w-z)\,
\rho_{\mathrm{l}}\left[\gamma_{\mathrm{l}}(i)\,\left({\frac{z}
{\zeta_{\mathrm{l},0}}}\right)^{n_{\mathrm{l}}}\right]}
{\int_{-t}^{t} \mathrm{d}w
\,\rho_{\mathrm{s}}\left[\gamma_{\mathrm{s}}(i)\,\left({\frac{w}
{\zeta_{\mathrm{s},0}}}\right)^{n_{\mathrm{s}}}\right]}},
\end{eqnarray}
where $t$ is the the LMC tidal radius.

We now integrate by part twice, a first time  obtaining:
\begin{eqnarray}\label{primaIntegrazionePerParti}
&\tau \le  {\frac{4\pi G}{c^{2}}}\int_{-t}^{t} \mathrm{d}w\,
\rho_{\mathrm{s}}\left[\gamma_{\mathrm{s}}(i)\,
\left({\frac{w}{\zeta_{\mathrm{s},0}}}\right)^{n_{\mathrm{s}}}\right]
\cdot\nonumber\\
&\cdot{\frac{\int_{-d}^{w}\,\mathrm{d}z\,
\int_{-d}^{z}\mathrm{d}u\,\rho_{l}\left[\gamma_{\mathrm{l}}(i)\,
\left({\frac{u}{\zeta_{\mathrm{l},0}}}\right)^{n_{\mathrm{l}}}\right]}
{\int_{-t}^{t}
\mathrm{d}w\rho_{\mathrm{s}}\,\left[\gamma_{\mathrm{s}}
(i)\,\left({\frac{w}{\zeta_{\mathrm{s},0}}}\right)^{n_{\mathrm{s}}}\right]}}
\end{eqnarray}
and a second time obtaining:
\begin{eqnarray}
&\tau \le  &{\frac{4\pi G}{c^{2}}}
\int_{-d}^{t}\mathrm{d}z\,\rho_{\mathrm{l}}
\left[\gamma_{\mathrm{l}}(i)\,\left({\frac{z}{\zeta_{\mathrm{l},0}}}
\right)^{n_{\mathrm{l}}}\right]
\,\cdot\nonumber \\
&&\cdot \left\{ \int_{-t}^{t}\,\mathrm{d}w \,{\frac{\int_{-d}^{w}
\mathrm{d}z \,
\rho_{\mathrm{l}}\left[\gamma_{\mathrm{l}}(i)\,\left({\frac{z}
{\zeta_{\mathrm{l},0}}}\right)^{n_{\mathrm{l}}}\right]}
{\int_{-d}^{t}\mathrm{d}z\,\rho_{\mathrm{l}}
\left[\gamma_{\mathrm{l}}(i)\,\left({\frac{z}{\zeta_{\mathrm{l},0}}}
\right)^{n_{\mathrm{l}}}\right] }}
\right.\cdot \nonumber  \\
&&\qquad \left. \left(
1-{\frac{\int_{-t}^{w}\mathrm{d}z\,\rho_{\mathrm{s}}
\left[\gamma_{\mathrm{s}}(i)\,\left({\frac{z}{\zeta_{\mathrm{s},0}}}
\right)^{n_{\mathrm{s}}}\right] }
{\int_{-t}^{t}\,\rho_{\mathrm{s}}\left[\gamma_{\mathrm{s}}(i)\,
\left({\frac{z}{\zeta_{\mathrm{s},0}}}\right)^{n_{\mathrm{s}}}\right]
\,\mathrm{d}z}} \right) \right\}.
\end{eqnarray}
In case of self lensing $\rho_{\mathrm{s}}$ and
$\rho_{\mathrm{l}}$ coincide. Moreover, for distances higher than
the tidal radius of the LMC, the star mass density  $\rho_{\mathrm{s}}
= 0$, and we can move the lower integration limit from $-t$ or
$-d$ to $-\infty$, and the upper from $t$ to $+\infty$. We  are
considering a line of view passing through the center of the LMC,
therefore we can use the symmetry property of the mass density
distributions with respect to change of sign of the parameter $z$
and we obtain:
\begin{eqnarray}
\tau \le & {\frac{\pi G}{c^{2}}}
\int_{-\infty}^{\infty}\mathrm{d}z\,\rho_{\mathrm{s}}
\left[\gamma_{\mathrm{s}}(i)\,\left({\frac{z}{\zeta_{\mathrm{s},0}}}
\right)^{n_{\mathrm{s}}}\right] \,\cdot \nonumber  \\& \cdot
\int_{-\infty}^{\infty}\mathrm{d}v \, \left[ 1-F_{s}^{2}(v)
\right],
\end{eqnarray}
where
\begin{equation}
F_{\mathrm{s}}(v) =
{\frac{2\int_{0}^{v}\mathrm{d}z\,\rho_{\mathrm{s}}
\left[\gamma_{\mathrm{s}}(i)\,\left({\frac{z}{\zeta_{\mathrm{s},0}}}
\right)^{n_{\mathrm{s}}}\right] }
{\int_{-\infty}^{\infty}\mathrm{d}z\,\rho_{\mathrm{s}}
\left[\gamma_{\mathrm{s}}(i)\,\left({\frac{z}{\zeta_{\mathrm{s},0}}}
\right)^{n_{\mathrm{s}}}\right] }}.
\end{equation}
A suitable change of variables gives:
\begin{eqnarray}\label{GouldIneqImprov}
\tau &\le  &{\frac{2\, \pi\,
G\,\zeta_{\mathrm{s},0}^{2}}{c^{2}}}
\left[\gamma_{\mathrm{s}}(i)\right]^{-{\frac{2}{n_{\mathrm{s}}}}}\cdot\nonumber\\
&&\int_{-\infty}^{\infty}\,\rho_{\mathrm{s}}\left(u^{n_{\mathrm{s}}}\right)
\,\mathrm{d}u \, \int_{0}^{\infty} \left[ 1-F_{\mathrm{s}}^{2}(v)
\right]\,\mathrm{d}v,
\end{eqnarray}
where
\begin{equation}
F_{\mathrm{s}}(v) =
{\frac{2\int_{0}^{v}\,\rho_{\mathrm{s}}\left(u^{n_{\mathrm{s}}}\right)
\,\mathrm{d}u}
{\int_{-\infty}^{\infty}\,\rho_{\mathrm{s}}\left(u^{n_{\mathrm{s}}}\right)
\,\mathrm{d}u}}
\end{equation}
and the integrations are now made on adimensional variables.

Let us observe that the inequality (\ref{GouldIneqImprov}) can be
applied for any inclination angle $i$ of the disk plane, not only
for $i << \pi/2$, as in \citet{gould95}, namely, we have no
divergence problem for $i\rightarrow \pi/2$. For instance, in the
case of a double exponential disk, with scale lengths respectively
equal to $R_{\mathrm{d}}$ and $\zeta_{\mathrm{d}}$, we obtain:
\begin{eqnarray}
\tau \le  &{\frac{2 \,\pi \, G
\,\zeta_{\mathrm{d}}^{2}}{c^{2}}}{\frac{1}
{\left[{\cos(i)}+{\frac{\zeta_{\mathrm{d}}}{R_{\mathrm{d}}}}\sin(i)\right]^{2}}}\,
\int_{-\infty}^{\infty}\,\rho_{\mathrm{s}}\left(u\right)
\,\mathrm{d}u \,\cdot  \nonumber\\ &\int_{0}^{\infty}
\left[1-F_{\mathrm{s}}^{2}(v)\right]\,\mathrm{d}v.
\end{eqnarray}
Or in the case of a gaussian profile of the mass density
distribution for the LMC bar, as in \citet{gyuk00}, with scale
length $\eta_{\mathrm{b}}$ along the axis bar and
$\zeta_{\mathrm{b}}$ in the orthogonal section, we obtain:
\begin{eqnarray}
\tau \le  &{\frac{2 \,\pi \,G
\,\zeta_{\mathrm{b}}^{2}}{c^{2}}}{\frac{1}{{\cos^{2}(i)}+
\left({\frac{\zeta_{\mathrm{b}}}{\eta_{\mathrm{b}}}}\right)^{2}\sin^{2}(i)}}\,
\int_{-\infty}^{\infty}\,\rho_{\mathrm{s}}\left(u^{2}\right)
\,\mathrm{d}u \,\cdot  \nonumber\\ &\int_{0}^{\infty}
\left[1-F_{\mathrm{s}}^{2}(v)\right]\,\mathrm{d}v.
\end{eqnarray}
For a bar having boxy contours and sharp edges as in the paper of
\citet{zhao00}, parameterized by an exponential profile with a
fourth degree polynomial at the exponent, with scale length
$\eta_{\mathrm{b}}$ along the axis bar and $\zeta_{\mathrm{b}}$ in
the orthogonal section, we obtain:
\begin{eqnarray}
\tau \le  &{\frac{2 \,\pi \,G
\,\zeta_{\mathrm{b}}^{2}}{c^{2}}}{\frac{1}{\sqrt{{\cos^{4}(i)}+
\left({\frac{\zeta_{\mathrm{b}}}{\eta_{\mathrm{b}}}}\right)^{4}\sin^{4}(i)}}}\,
\int_{-\infty}^{\infty}\,\rho_{\mathrm{s}}\left(u^{4}\right)
\,\mathrm{d}u \,\cdot  \nonumber\\ &\int_{0}^{\infty}
\left[1-F_{\mathrm{s}}^{2}(v)\right]\,\mathrm{d}v.
\end{eqnarray}
Eq. (\ref{GouldIneqImprov}) is the improved version of Eq. (2.8)
of \citet{gould95}.

Proceeding in the same way as in \citet{gould95}, we obtain the
second Gould inequality, relating the optical depth with the
mass-weighted velocity dispersion.
\begin{equation}\label{secondGouldIneqImprov}
\tau \le 2{\frac{<v^{2}>}{c^{2}}}
\left[\gamma_{\mathrm{s}}(i)\right]^{-{\frac{2}{n_{\mathrm{s}}}}}.
\end{equation}

Let us note that these inequalities are based on two
properties:
\begin{itemize}
\item{} that the star density  be a
function of an {\it homogeneous polynomial of the variables}, of any degree;
\item{} that the line of view pass through the center, in such a
way we can use the symmetry by reflection of the mass density
distributions.
\end{itemize}

Let us note, moreover, that the inequalities
(\ref{GouldIneqImprov}) and (\ref{secondGouldIneqImprov}) can not
be applied to the model of the LMC disk and bar we assumed in this
paper, because the first property, that of homogeneity, is
lacking.

It is therefore useful to derive a more general inequality,
applicable to all kinds of density distribution.

%
\subsection{A more general inequality}
We start from equation (\ref{primaIntegrazionePerParti}), but we
relax the hypothesis that  the mass density distribution be a
function of an homogeneous polynomial:
\begin{equation}\label{startForGeneralInequality}
\tau \le  {\frac{4\pi G}{c^{2}}}{\frac{\int_{-t}^{t} \mathrm{d}w\,
\rho_{\mathrm{s}}(w) \int_{-d}^{w}\mathrm{d}z\,
\int_{-d}^{z}\mathrm{d}u\,\rho_{\mathrm{l}}(u)}
{\int_{-t}^{t}\mathrm{d}w\,\rho_{\mathrm{s}}\,(w)}}.
\end{equation}

Let us divide in three part the triangular region of integration
in the plane $(u,z)$, delimited by the bisector of the first and
third quadrant, by the parallel to the $u$-axis of equation $z =
w$ and by the parallel to $z$--axis of equation $u = -d$: the
first region is constituted by the triangle delimited by the
bisector, the line $u = -d$ and the $u$--axis; the second by the
rectangle delimited by the $u$--axis, the $z$--axis and the lines
of equation $z=w$ and $u = -d$; the third by the triangle
delimited by the bisector, the z-axis and the line of equation $z
= w$. Let us remember that $w$ can assume any value between $-t$
and $t$. In this way, the right hand side of Eq.
(\ref{startForGeneralInequality}) is given by the sum of three
terms:
\begin{eqnarray}
&{\frac{4 \,\pi \, G}{c^{2}}}\left\{
\int_{-d}^{0}\mathrm{d}z\,\int_{-d}^{z}\mathrm{d}u\,\,\rho_{\mathrm{l}}(u)
\, +\,\int_{-d}^{0}\mathrm{d}z\,\rho_{\mathrm{l}}(z)
{\frac{\int_{-t}^{t} \mathrm{d}w\,
w\,\rho_{\mathrm{s}}(w)}{\int_{-t}^{t}
\mathrm{d}w\, \rho_{\mathrm{s}}(w)}}\right. \nonumber\\
&\left. +\,{\frac{\int_{-t}^{t} \mathrm{d}w\, \rho_{\mathrm{s}}(w)
\int_{0}^{w}\mathrm{d}z\,
\int_{0}^{z}\mathrm{d}u\,\rho_{\mathrm{l}}(u)}
{\int_{-t}^{t}d\mathrm{w}\,\rho_{\mathrm{s}}\,(w)}}\right\}.
\end{eqnarray}
After an integration by part, the first term becomes
$${\frac{4 \,\pi \,
G}{c^{2}}}\int_{0}^{d}\mathrm{d}z\,z\,\rho_{\mathrm{s}}(-z);
$$
the second term is null, thanks  to the symmetry by reflection of
$\rho_{\mathrm{s}}(z)$. As regard the third term we observe that
for any $z$ belonging to the interval $(-t,t)$ it  always results
$$
\int_{0}^{z}\mathrm{d}u\,\rho_{\mathrm{l}}(u) \le
\int_{0}^{t}\mathrm{d}u\,\rho_{\mathrm{l}}(u)
$$
and therefore
\begin{eqnarray}
{\frac{\int_{-t}^{t} \mathrm{d}w\, \rho_{\mathrm{s}}(w)
\int_{0}^{w}\mathrm{d}z\,
\int_{0}^{z}\mathrm{d}u\,\rho_{\mathrm{l}(}u)}
{\int_{-t}^{t}\mathrm{d}w\,\rho_{\mathrm{s}}\,(w)}}\le
&\int_{-t}^{t}
\mathrm{d}w\,w\, \rho_{\mathrm{s}}(w)\,\cdot \nonumber\\
&{\frac{ \int_{0}^{t}\mathrm{d}u\,\rho_{\mathrm{l}}(u)}
{\int_{-t}^{t}\mathrm{d}w\,\rho_{\mathrm{s}}\,(w)}}.
\end{eqnarray}
This way we obtain that also the third term is less than zero.
In conclusion we obtain the searched inequality for the optical
depth, valid for any mass density distribution of lenses. In
particular for lenses in the galactic halo we obtain:
\begin{equation}\label{generalInequalityGH}
\tau_{\mathrm{GH}} \le  {\frac{4\,\pi\,
G}{c^{2}}}\int_{0}^{d}\mathrm{d}z\,z\,\rho_{\mathrm{l,GH}}(-z),
\end{equation}
for lenses in the LMC halo we obtain
\begin{equation}\label{generalInequalityLMCH}
\tau_{\mathrm{LMCH}} \le  {\frac{4\,\pi\,
G}{c^{2}}}\int_{0}^{t}\mathrm{d}z\,z\,\rho_{\mathrm{l,LMCH}}(z),
\end{equation}
and for self lensing we obtain
\begin{equation}\label{generalInequalitySL}
\tau_{\mathrm{SL}} \le  {\frac{4\,\pi\,
G}{c^{2}}}\int_{0}^{t}\mathrm{d}z\,z\,\rho_{\mathrm{s}}(z).
\end{equation}
Applying this last inequality to the LMC mass density distribution
used in our (coplanar) model we find
$$
\tau_{\mathrm{SL}} \le 6 \times 10^{-8},
$$
only about 20\% higher than the calculated value.

Notice that on the right hand side of the
inequality (\ref{generalInequalityGH})
$
\rho_{\mathrm{l,GH}}(-z)\ne \rho_{\mathrm{l,GH}}(z),
$
since the galactic halo density is not symmetric with respect to the
LMC center. The inequality, therefore, cannot be obtained by a
trivial dropping of the factor $(1-z)$ in the integrand of the
expression defining the galactic halo optical depth.

\section{Summary} \label{sec:conclusioni}

The great interest in the \emph{location} of the observed
microlensing events towards the LMC is motivated by the need to
give an answer to the question of their \emph{nature}. Namely,
whether (or not) all the events can be attributed to known
(luminous) populations, so to exclude (or not) the possibility for
a dark component in the halo in the form of MACHOs.

In this paper we are mainly concerned with the possible
self--lensing origin of the observed microlensing events. In
particular we have considered the results of the MACHO survey. We
use the recent picture of the LMC disk given by \citet{marel02}, and
we explore different geometries for the bar component, as well as
a reasonable range for the velocity dispersion for the bar
population.

One interesting feature, essentially linked to the assumed
disk geometry, is an evident near--far asymmetry of the optical
depth for lenses located in the LMC Halo (this is not
expected, with the possible exception of the inner region, for the
self--lensing population). Indeed, similarly to the case of M31
\citep{crotts,jetzer}, and as first pointed out by
\citet{gould93}, since the LMC disk is inclined, the optical depth
is higher along lines of sight passing through larger portions of
the LMC halo. We show that such a spatial asymmetry, beyond the
one expected from the observational strategy alone, is indeed
present in the observed events. With the care suggested by the
small number of detected events on which this analysis is based,
this can be looked at, as yet observed by \citet{gould93}, as a
signature of the presence of an extended halo around the LMC.

In the central region the microlensing signatures are strongly
dependent on the assumed bar geometry. In particular, we have
studied the variation (that can be as large as 50\%) in the
self--lensing optical depth due to the different geometry of the
bar. However, the available data do not allow us to meaningfully
explore in more detail this aspect.

As a further step in our analysis, we have studied the
microlensing rate. Keeping in mind \citet{evans00} observation
that the timescale distribution of the events and their spatial
variation across the LMC disk offers possibilities of identifying
the dominant lens population, we have carefully characterized the
ensemble of observed events under the hypothesis that all of them
do belong to the self--lensing population. Through this analysis
we have been able to identify a large subset of events that can
not be accounted as part of this population. The introduction
of a non coplanar bar component with respect to the disk turns out to enhance this result.
Again, the small amount of events at disposal does not yet allow
us to draw sharp conclusions, although, the various arguments
mentioned above are all consistent among them and converge quite
clearly in the direction of excluding self lensing as being the
major cause for the events.

Once more observations will be available, as will hopefully be the
case with the SuperMacho experiment under way \citep{stubbs}, the
use of the above outlined methods can bring to a definitive answer
to the problem of the location of the MACHOs and thus also to
their nature.

\begin{acknowledgements}
The authors wish to thank the anonymous referee for his comments 
which improved the quality of this work and Chiara Mastropietro for useful
discussion on the LMC morphology. LM and
SCN are partially supported by the Swiss National Science
Foundation and SCN is also partially supported by the Tomalla
Foundation. GS wishes to thank the Institute of Theoretical
Physics of the University of Z\"urich for the kind hospitality.
\end{acknowledgements}

------------------------------------------------------------------

\bibliographystyle{aa}

\end{document}